\documentclass[journal,12pt,draftclsnofoot,onecolumn]{IEEEtran}

\usepackage[usenames,dvipsnames]{color}
\usepackage{setspace}
\usepackage{amsmath}
\usepackage{graphicx}
\usepackage{epstopdf}
\usepackage{amssymb}
\usepackage[hyphens]{url}
\usepackage{cite}
\usepackage{subfig}
\usepackage[usenames,dvipsnames]{color}
\usepackage{setspace}
\usepackage{bm}
\usepackage[utf8]{inputenc}
\usepackage[english]{babel}
\usepackage{amsthm}
\usepackage{enumerate}
\usepackage{siunitx}

\theoremstyle{remark}
\newtheorem{remark}{Remark}

\DeclareMathOperator*{\argmaxA}{arg\,max}

\newcommand{\comm}{\mathrm{com}}
\newcommand{\cm}{\mathrm{c}}

\newcommand{\cefm}{\mathrm{CEF}}

\newcommand{\Es} {{\mathcal{E}_{\mathrm{s}}} }

\newcommand{\Erm}{\scnr}

\newcommand{\e}[1]{{\mathbb E}\left[ #1 \right]}
\newcommand{\me}{\mathrm{e}}

\newcommand{\Ga}{\mathbf{a}_{\mathrm{128}}}
\newcommand{\Gb}{\mathbf{b}_{\mathrm{128}}}
\newcommand{\Gu}{\mathbf{a}_{\mathrm{512}}}
\newcommand{\Gv}{\mathbf{b}_{\mathrm{512}}}

\newcommand{\hum}{h_\um}

\newcommand{\isim}{\mathrm{ISI}}
\newcommand{\jm}{\mathrm{j}}

\newcommand{\mr}{\mathrm{rad}}

\newcommand{\mapm}{\mathrm{map}}

\newcommand{\ND}{N_{\mathrm{D}}}

\newcommand{\Nl}{N_\mathrm{p}}

\newcommand{\NT}{P}
\newcommand{\Np}{P_{\mathrm{r}}}
\newcommand{\Nt}{N_{\txm}}
\newcommand{\Nr}{N_\mrx}

\newcommand{\nm}{\mathrm{n}}

\newcommand{\Pt}{P}

\newcommand{\psm}{\mathrm{g}}

\newcommand{\Rj}{R_1}

\newcommand{\mrx}{\mathrm{RX}}

\newcommand{\scnr}{\mathrm{\zeta}_{\mr}}
\newcommand{\snr}{\mathrm{\zeta}_{\comm}[m]}
\newcommand{\TD}{T_{\mathrm{D}}}
\newcommand{\Ts}{T_{\mathrm{s}}}
\newcommand{\Tc}{T_{\mathrm{s}}}

\newcommand{\trm}{\mathrm{tr}}
\newcommand{\txm}{\mathrm{TX}}

\newcommand{\thm}{\mathrm{th}}

\newcommand{\taud}{\tau_\mathrm{d}}
\newcommand{\tm}{\mathrm{t}}

\newcommand{\um}{0}

\newcommand{\wt}{\mathbf{f}_\txm}

\pdfoutput=1

\title{IEEE 802.11ad-based Radar: An Approach to Joint Vehicular Communication-Radar System}
\author{{Preeti Kumari, Junil Choi, Nuria Gonz\'alez-Prelcic, and Robert W. Heath Jr.}
\thanks{ Preeti Kumari and Robert W. Heath Jr. are with the Wireless Networking and Communications Group, the University of Texas at Austin, TX 78712-1687, USA (e-mail: \{preeti\_kumari, rheath\}@utexas.edu).
J. Choi is with the Department of Electrical Engineering, POSTECH, Pohang, Gyeongbuk, Korea 37673 (e-mail:junil@postech.ac.kr).
Nuria Gonz\'alez-Prelcic is with the Department of Signal Theory and Communications, Universidade de Vigo, Vigo, Spain 36310 (email: nuria@gts.uvigo.es)
This research was partially supported by the U.S. Department of Transportation through the Data-Supported Transportation Operations and Planning (D-STOP) Tier 1 University Transportation Center and by the Texas Department of Transportation under Project 0-6877 entitled “Communications and Radar-Supported Transportation Operations and Planning (CAR-STOP)”. This work was also supported by a gift from National Instruments.
}}

\begin{document}
\maketitle
\begin{abstract}
Millimeter-wave (mmWave) radar is widely used in vehicles for applications such as adaptive cruise control and collision avoidance. In this paper, we propose an IEEE 802.11ad-based radar for long-range radar (LRR) applications at the 60 GHz unlicensed band. We exploit the preamble of a single-carrier (SC) physical layer (PHY) frame, which consists of Golay complementary sequences with good correlation properties, as a radar waveform. This system enables a joint waveform for automotive radar and a potential mmWave vehicular communication system based on IEEE 802.11ad, allowing hardware reuse. To formulate an integrated framework of vehicle-to-vehicle (V2V) communication and LRR based on a mmWave consumer wireless local area network (WLAN) standard, we make typical assumptions for LRR applications and incorporate the full duplex radar assumption due to the possibility of sufficient isolation and self-interference cancellation. We develop single- and multi-frame radar receiver algorithms for target detection as well as range and velocity estimation within a coherent processing interval. Our proposed radar processing algorithms leverage channel estimation and time-frequency synchronization techniques used in a conventional IEEE 802.11ad receiver with minimal modifications. Analysis and simulations show that in a single target scenario, a Gbps data rate is achieved simultaneously with cm-level range accuracy and cm/s-level velocity accuracy. The target vehicle is detected with a high probability of detection ($>$99.9$\%$) at a low false alarm of 10$^{-6}$ for an equivalent isotropically radiated power (EIRP) of 43 dBm up to a vehicle separation distance of 200 m.
\end{abstract}
\section{Introduction}
Vehicular radar and communication are the two primary means of using radio frequency (RF) signals in transportation systems. Automotive radars provide high-resolution sensing using proprietary waveforms in the mmWave band \cite[Ch. 4]{saponara2014highly}. They enable safety- and comfort-related functions, such as adaptive cruise control, blind spot warning, and pre-crash applications \cite{hasch2012millimeter}. Vehicular communication allows vehicles to exchange safety messages or raw sensor data for applications such as forward collision warning, do-not-pass warning, and cooperative adaptive cruise control \cite{papadimitratos2009vehicular}. The default vehicular communication standard is dedicated short-range communication (DSRC), which is designed for low-latency using a WLAN-based physical layer and is allocated 75 MHz of licensed spectrum in the 5.9 GHz band \cite{kenney2011dedicated}. Unfortunately, DSRC achieves data rates of at most 27 Mbps, much less than the requirement for applications such as full automated driving (based on raw sensor data exchange to enlarge sensing range), or precise navigation (based on downloading high-definition 3D maps), which require Gbps data rates \cite{choi2016millimeter}. 

A solution to realize the next generation of high data rate connected vehicles is to exploit the large bandwidths available in the mmWave spectrum. This could be achieved using a 5G solution, a 60 GHz unlicensed solution, or a proprietary waveform in dedicated spectrum \cite{VaShiBan:Millimeter-Wave-Vehicular:16}. Additionally, it is not only interesting to achieve higher data rates in vehicular communications, but it is also beneficial to have a joint communication and radar system that allows hardware reuse. In the past half-decade, a number of joint communication-radar systems have been proposed (see, e.g., \cite{han2013joint} and the references therein). These approaches can be mainly classified as either a simultaneous system or a non-simultaneous system. In a simultaneous system, a single-carrier \cite{saddik2007ultra, sturm2011waveform} or a multi-carrier waveform \cite{berger2010signal, reichardt2012demonstrating, sturm2011waveform} are used for both communication and radar at the same time. In a non-simultaneous system, radar and communication operate in different time intervals \cite{zhang200724ghz,han2010radar}. Most of the prior work \cite{han2013joint,saddik2007ultra, sturm2011waveform,berger2010signal, zhang200724ghz,han2010radar} used waveforms that are not based on a communication standard.

OFDM waveforms are popular for implementing simultaneous joint communication-radar systems at sub-6 GHz frequencies\cite{berger2010signal, sturm2011waveform, reichardt2012demonstrating}. In \cite{sturm2011waveform}, the radar parameters are estimated by leveraging the channel estimation technique for OFDM communication systems, where the samples obtained at the output of the OFDM communication receiver before channel equalization is divided by the known transmitted data symbol to obtain the DFT of the channel coefficients. In \cite{berger2010signal}, radar parameters are estimated using classical correlation-based (matched filter) radar processing approaches that exploit OFDM baseband signals. The independence of the estimated channel coefficients from the transmitted data in \cite{sturm2011waveform} allows a higher dynamic range (between the strongest and the weakest reflection) as compared to \cite{berger2010signal}, without sacrificing processing gain and resolution. The sidelobe levels in \cite{sturm2011waveform}, however, are not ideal for radar ranging. In \cite{reichardt2012demonstrating}, the IEEE 802.11p V2V communication standard was analyzed as in \cite{sturm2011waveform} for automotive radars. The IEEE 802.11p-based radar, however, cannot achieve cm-level range and cm/s-level velocity resolution, which are desirable in automotive radars \cite{hasch2012millimeter}, due to insufficiently low bandwidth. OFDM-based simultaneous systems also suffer from a high peak-to-average power ratio (PAPR), unlike traditional automotive radars with frequency modulated continuous wave (FMCW) waveform that has PAPR of 0 dB. 

In this paper, we propose a mmWave joint vehicular communication and radar system. We build our approach around the IEEE 802.11ad mmWave WLAN standard, reusing the same waveform for automotive radar. This allows us to exploit the same spectrum and to leverage shared hardware based on the mmWave consumer WLAN standard. The approach is reasonable because the most prevalent vehicular communication standard, DSRC, is based on a WLAN standard. The use of a standard mmWave waveform, which provides access to a large bandwidth, will lead to significant advantages in terms of higher data rates for communication and better accuracy/resolution for radar operation compared with approaches based on sub-6 GHz frequencies. The integration of cooperative communication approaches and autonomous radar sensing solutions will improve the system performance in automotive safety and efficiency applications due to the mutual exchange of complementary information (e.g., enhancing the radar imaging accuracy \cite{Han:2016:ARC:2980100.2980106} or reducing beam training overhead for communications \cite{choi2016millimeter}). The joint system will also lead to a potential increase in the penetration rate of mmWave communication in vehicles and enhanced security \cite{YehChoPre:Security-in-automotive-radar:16}.  

In this initial study of IEEE 802.11ad-based radar, we make several typical assumptions for LRR applications: 1) a target vehicle can be represented by a single point model \cite{rohling1996cfar} and 2) the location, velocity,
and radar cross section of a target vehicle remain constant during a coherent processing interval (CPI) \cite{rohling2001waveform}. We also assume full-duplex radar operation due to sufficient isolation and self-interference cancellation provided by the spacing between the TX and the RX arrays, use of efficient circulators \cite{estep2014magnetic}, TX/RX beamforming \cite{LiJosTao:Feasibility-study-on-full-duplex:14}, and the possibility of further suppression in the digital, analog-circuit, or antenna domains \cite{sabharwal2013band}. These assumptions are further described in Section~\ref{sec:System}. 

The main contributions of this paper are summarized as follows.
\begin{itemize}
\item A system model is proposed for joint vehicle-to-vehicle communication and long-range radar using the SC PHY frame of IEEE 802.11ad. It captures the nuances of the channel description for both communication and radar systems along with the signal model for WLAN-based transmitter and receiver.

\item Single- and multi-frame radar algorithms are developed for single- and multi-target detection as well as range and velocity estimation. These algorithms exploit the IEEE 802.11ad preamble and leverage conventional WLAN time-frequency synchronization and channel estimation techniques per frame.

\item Simulations are performed to evaluate the performance of the proposed joint communication-radar system, which meets the required LRR specifications \cite{hasch2012millimeter,Continental}. In a single target scenario, we achieve a Gbps communication data rate simultaneously with the cm/s-level velocity accuracy using multiple frames (in a CPI $>$ 0.06 ms) and the cm-level range accuracy using a single frame. The velocity and range accuracy is measured quantitatively using root MSE. The target is detected with a high probability of detection ($>$ 99.9$\%$) at a significantly low false alarm rate of $10^{-6}$ up to a range of 200 m \cite{Continental}. In a multi-target situation, we can achieve a range resolution of $<$ 0.1 m and a velocity resolution of $<$ 0.6 m/s using multiple frames in a 4.2 ms CPI, which is less than CPI duration typically used for LRR processing (e.g., \cite{rohling2001waveform} uses a CPI of 10 ms). 
 
\item Theoretical performance analysis using the Cramer-Rao lower bound (CRLB) is provided for the single-frame target range and velocity estimation algorithms following the approach in \cite[Ch. 7]{richards2005fundamentals}. The CRLB is derived for the velocity estimation using multiple IEEE 802.11ad SC PHY frames in a single target scenario with additive Gaussian clutter-plus-noise. In numerical simulations, we achieve the velocity MSE very close to its CRLB. The single-frame range estimation MSE is quite close to its CRLB and the slight difference between them, which is less than 2 cm$^2$, is due to the limited accuracy of the employed WLAN symbol synchronization techniques \cite{liuall,preyss2015digital}.
\end{itemize} 

Our previous work in \cite{preeti2015} was the first to propose the idea of using IEEE 802.11ad for a joint vehicular communication and radar system. There were some limitations in \cite{preeti2015} that are overcome in this paper. First, the system model was developed only for a single target model using a single frame. It did not include a multi-target model or a false alarm rate detection performance metric. Second, the Doppler shift estimation was not accurate at low and medium signal-to-noise ratio (SNR). Third, it did not provide a theoretical insight into the performance. Our new work overcomes these limitations and provides a further in-depth analysis and simulation of the proposed IEEE 802.11ad-based communication-radar system.

The rest of the paper is organized as follows. A summary of the preamble sequences for an SC PHY frame of IEEE 802.11ad is presented in Section II. In Section III, an integrated system model of LRR and V2V communication is developed. Section IV proposes different single- and multi-frame processing techniques and analyzes their theoretical performance for radar parameter estimation. Numerical results and performance evaluations are described in Section V, while the conclusion follows in Section VI. 

\textbf{Notation:} We use the following notation throughout the paper:  vectors are denoted by boldface lowercase letters $\mathbf{a}$, matrices by boldface capital letters $\mathbf{A}$, and scalar values by $a$, $A$. The $n^{\mathrm{th}}$ component of vector $\mathbf{a}$ is written as $a[n]$ and the $(\ell,m)^{\mathrm{th}}$ element of matrix $\mathbf{A}$ is denoted by $A[\ell,m]$. We use the notation $\vert {c} \vert $ for the magnitude of ${c}$, $\angle {c}$ for the phase of ${c}$, and $a(t) \ast b(t)$ for the convolution between signals $a(t)$ and $b(t)$. $\vert \vert \mathbf{B} \vert \vert _{\mathrm{F}}$ is the Frobenius norm, $\mathbf{B}^{\mathrm{*}}$ is the conjugate transpose, $\mathbf{B}^{\mathrm{T}}$ is the transpose, and $\mathbf{B}^{\mathrm{c}}$ is the conjugate of matrix $\mathbf{B}$. We use the notation $\mathcal{N_C}(\mu,\sigma^2)$ to denote a complex circularly symmetric Gaussian random variable with mean $\mu$ and variance $\sigma^2$. The subscript $\mr$ refers to radar, $\comm$ refers to communication, $\txm$ refers to a transmitter, $\mrx$ refers to a receiver, $\cm$ refers to clutter, and $\mathrm{n}$ refers to noise. Frequently used symbols in the paper are summarized in Table~\ref{table_Symbols}.
 \begin{table}[!t] 
\renewcommand{\arraystretch}{1.3}
 \caption{Frequently Used Symbols}
\label{table_Symbols}
\centering
\begin{tabular}{|c|l|}
\hline
Notation & Description \\
\hline
$\Ts$ & Symbol period\\
\hline
$\Es$ & Signal energy per symbol at the transmitter\\
\hline
$T$ & CPI duration \\
\hline
$M$ & Number of frames in a CPI \\
\hline
$\tau_{p}$ & Round-trip delay of the $p^\thm$ target\\
\hline 
$\rho_p$ & Range of the $p^\thm$ target\\
\hline
$\nu_{p}$ & Doppler shift of the $p^\thm$ target\\
\hline
$v_p$ & Relative radial velocity of the $p^\thm$ target\\
\hline
$\scnr$ & SCNR of the received radar signal \\
\hline 
\end{tabular}
\end{table}
\section{The IEEE 802.11ad Preamble}
In this section, we review the preamble of the IEEE 802.11ad SC PHY frame and compute its ambiguity function to assess its suitability as an automotive radar waveform for single- and multi-target vehicular scenarios. 

\subsection{Frame Structure}
\begin{figure}[!t]
\centering
\includegraphics[scale = 0.75]{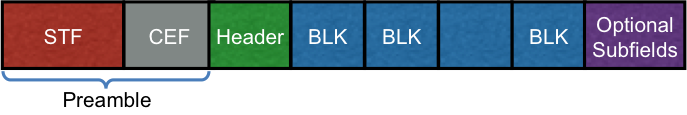}
 \caption{The structure of the SC PHY IEEE 802.11ad frame, which consists of a preamble, a header, communication data blocks (BLKs) and optional beam training fields. The preamble has many repeated sequences with good correlation properties that makes it suitable for radar.}
\label{fig_frame}
  \vspace{-0.45cm}
\end{figure}

\begin{figure}[!t]
\centering
\includegraphics[scale = 0.8]{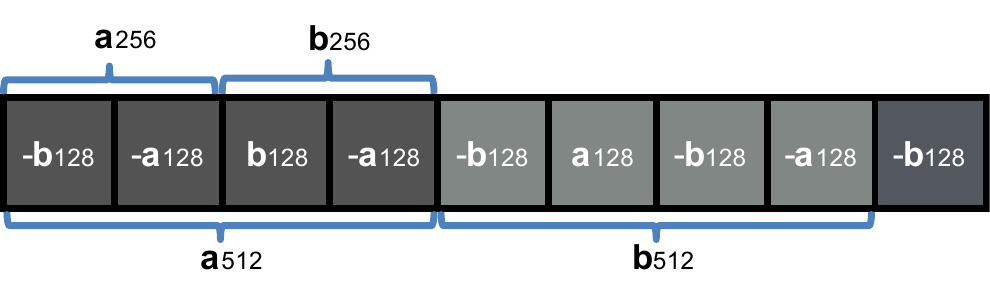}
 \caption{Extracted CEF for an SC PHY frame. It contains a 128 sample Golay complementary pair, denoted by $[\Ga \; \Gb]$, a 256 sample Golay complementary pair, denoted by $[\mathbf{a}_{256} \; \mathbf{b}_{256}]$, and a 512 sample Golay complementary pair, denoted by $[\Gu \; \Gv]$.}
\label{fig_CE}
  \vspace{-0.45cm}
\end{figure}
An IEEE 802.11ad SC PHY frame structure is shown in Fig.~\ref{fig_frame}. In this paper, we exploit the preamble, which is composed of the short training field (STF) and the channel estimation field (CEF). The preamble in the SC PHY frame is similar to that in other PHY frames of IEEE 802.11ad \cite{ieee2012wireless}. Therefore, the findings using SC PHY modulation can be extended to other PHY preambles.

The STF is composed of sixteen repeated 128 sample Golay complementary sequence, $\Ga$, followed by its binary complement $-\Ga$ \cite{ieee2012wireless}. It is used in communication for frame synchronization and frequency offset estimation. The frame synchronization algorithm can be leveraged for range estimation and the frequency offset estimation technique can be used for velocity estimation of a radar target, as explained in Section \ref{Sec:ProRx}.


The CEF consists of a 512 sample Golay complementary pair, denoted by $[\Gu \; \Gv]$ and is followed by $-\Gb$, as shown in Fig.~\ref{fig_CE}. It is used to estimate the communication channel parameters. The channel estimation algorithm can also be leveraged for target range and velocity estimation. In this paper, we propose radar algorithms for target detection as well as range and velocity estimation by using both the STF and the CEF, either jointly or by using the CEF after the STF, as explained in Section \ref{Sec:ProRx}. The algorithms that exploit the STF and the CEF jointly can be used for a longer range of operation as compared to the one that uses the CEF after the STF. The algorithms that use the CEF after the STF, however, leverage the perfect auto-correlation property of Golay complementary sequences desirable in automotive radars.
\subsection{Ambiguity Function}
\begin{figure}[!t]
\centering
\includegraphics[scale = 0.35]{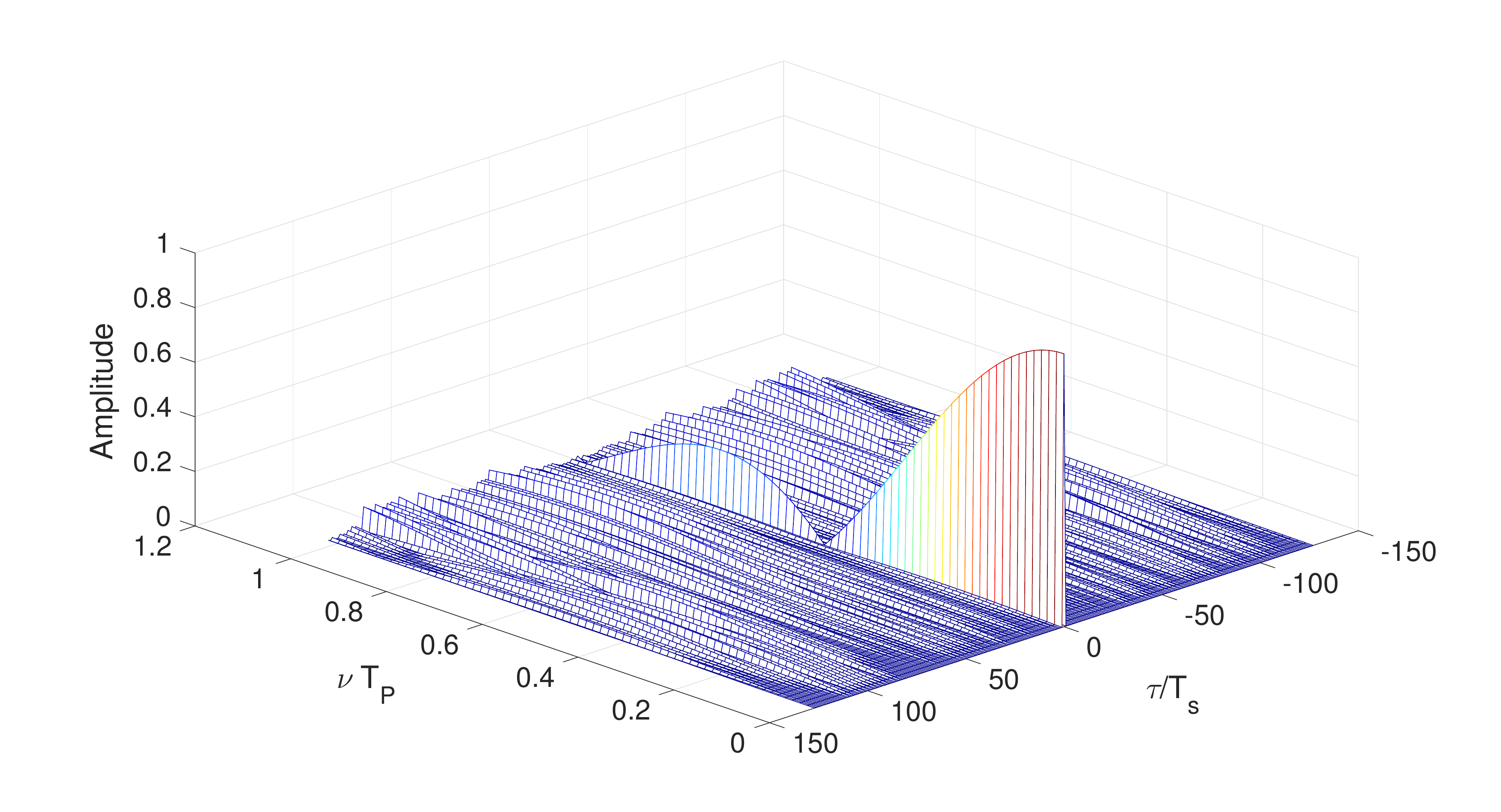}
 \caption{Ambiguity function diagram of the 512 sample Golay complementary pair with duration $T_\mathrm{p}$. Here, $\tau$ denotes delay,  $ \Ts$ represents symbol period, $\nu$ denotes Doppler shift, and $T_\mathrm{p} = 512 \Ts$.}
 \label{fig_compAmb}
  \vspace{-0.45cm}
\end{figure}

To establish the suitability of the IEEE 802.11ad preamble for automotive radar, we use the ambiguity function. The ambiguity function diagram of $[\Gu \; \Gv]$ is computed using the closed form solution in \cite{turyn1963ambiguity} and is shown in Fig.~\ref{fig_compAmb}. 
The zero-Doppler cut of the ambiguity function indicates that $[\Gu \; \Gv]$ has a perfect auto-correlation with no sidelobe along the zero-Doppler axis. This characteristic makes it ideal for radar target detection, which does not exist in FMCW signals typically used in LRR \cite{bazzi2012estimation}. The ambiguity function of $[\Gu \; \Gv]$ also depicts that it is less tolerant to large Doppler shifts. These sequences, however, are still acceptable for LRR due to the small normalized Doppler shift inherent in the vehicular environment. 

\section{System Model} \label{sec:System}
 \begin{figure}[!t]
\centering
\includegraphics[scale=0.6]{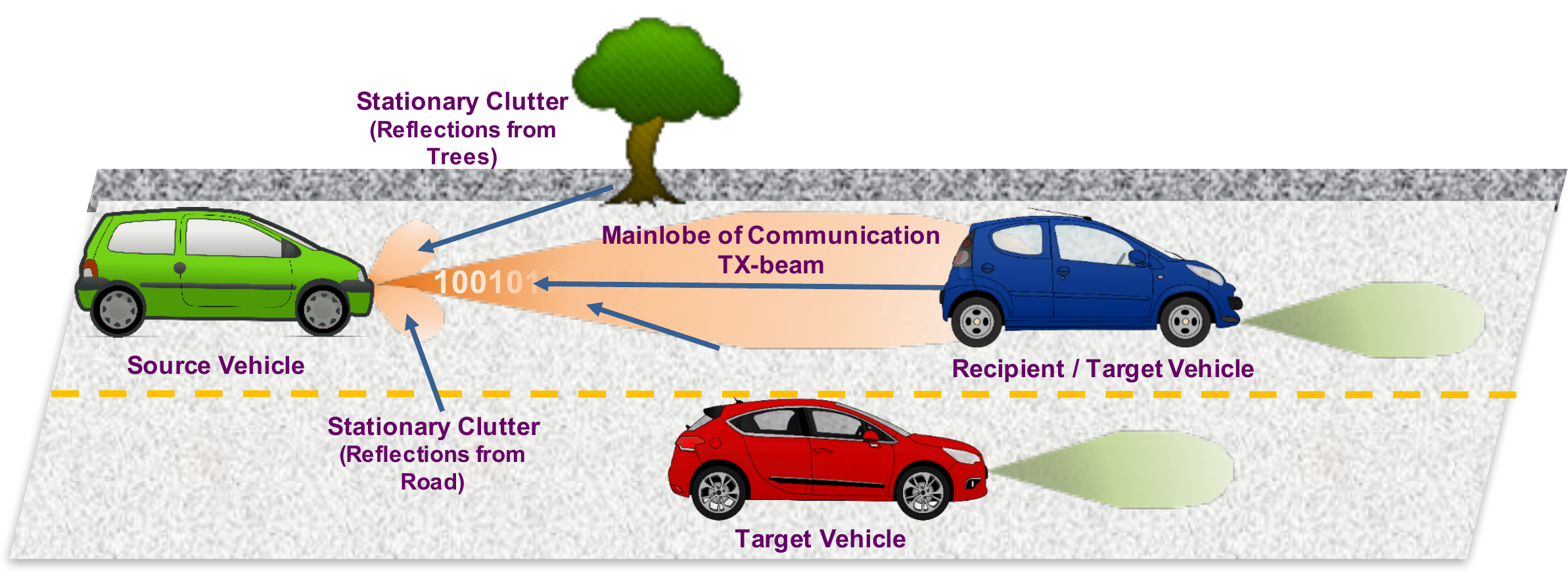}
 \caption{Illustration of a vehicular scenario, where a source vehicle transmits an IEEE 802.11ad signal to a recipient vehicle receiver and uses the echoes from target vehicles and clutter to derive target range and velocity estimates at the IEEE 802.11ad-based radar receiver mounted on the source vehicle.}
\label{fig_scenario}
 \vspace{-0.45cm}
\end{figure}
In this section, we formulate the signal and channel models for the proposed vehicular communication and automotive radar system.
We consider the use case where a source vehicle sends an IEEE 802.11ad waveform to a recipient vehicle receiver and uses the echoes from a single or multiple target(s) to derive range and velocity estimates, as shown in Fig.~\ref{fig_scenario}. We assume a multiple-antenna system for the joint communication-radar with an $N_\txm$-element transmit (TX) antenna array mounted on the source vehicle, and an $N_\mrx$-element receive (RX) antenna array mounted on both the source and recipient vehicles. First, we develop the signal model for the IEEE 802.11ad waveform at the source vehicle, which serves as the TX signal for both communication and radar systems simultaneously. Second, we describe the one-way V2V communication channel and the two-way single and multi-target LRR channels at the mmWave band. Finally, we develop signal models for the communication receiver at the recipient vehicle and the radar receiver at the source vehicle.

\subsection{Transmit Signal Model}
The complex baseband continuous-time representation of the IEEE 802.11ad waveform is
\begin{equation} \label{eq:TXCont}
x(t) = \sqrt{{\mathcal{E}_{\mathrm{s}}}}  \sum_{n =-\infty}^{\infty}  s[n] g_\txm(t -n\Tc),
\end{equation}
where $\Es$ is the signal energy per symbol at the transmitter, $g_\txm(t)$ is the unit energy TX pulse-shaping filter, $\Ts$ is the sample duration, and $s[n]$ is the transmitted symbol sequence corresponding to a single-carrier waveform of IEEE 802.11ad normalized such that $\e{\vert s[n] \vert^2 } = 1$. The symbol period $ \Tc \approx 1/W$, where $W$ is the signaling bandwidth. The IEEE 802.11ad specification defines the RX filter for error vector magnitude (EVM) measurement as a root-raised cosine (RRC) filter with a roll-off factor of 0.25, but a specific TX pulse shaping is not specified. Therefore, in numerical simulations, we have assumed a unit energy RRC waveform with the same roll-off factor for the TX pulse shaping filter $g_\txm(t)$ and the RX pulse shaping filter $g_\mrx(t)$.

\begin{figure}[!t] 
\begin{minipage}[h]{\textwidth}
  \centering
  \centerline{\includegraphics[scale =0.55]{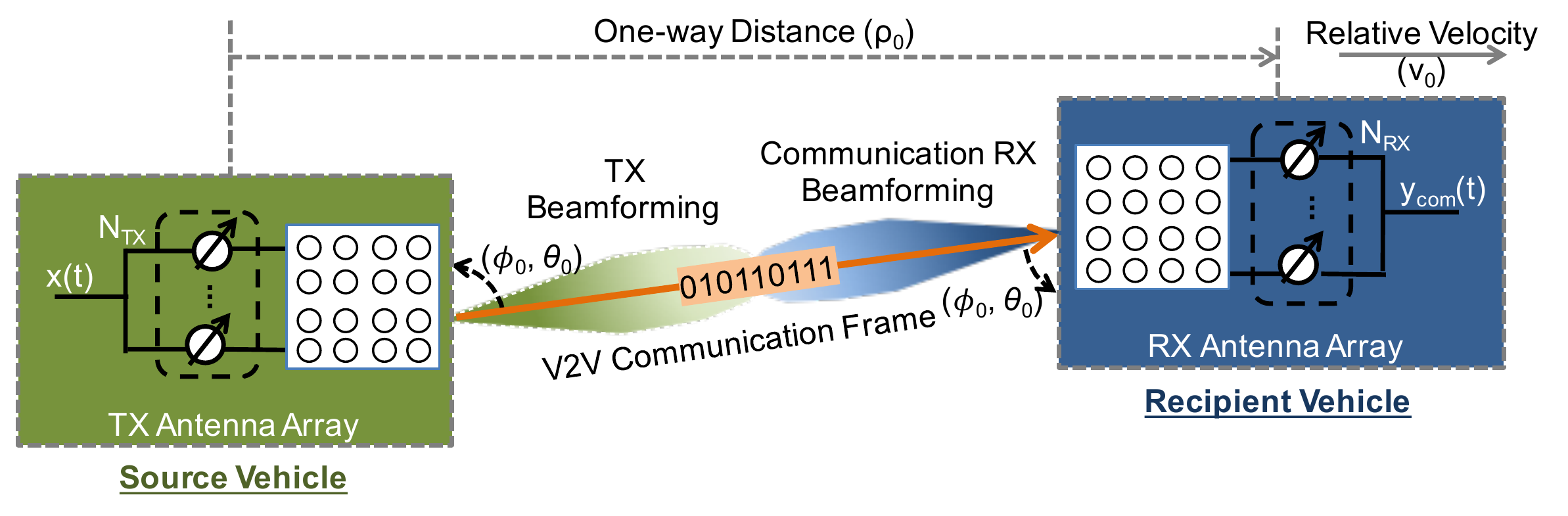}}
  {(a) One-way communication channel. Here, $y_\comm(t)$ denotes the continuous-time received communication signal.}\medskip
  \end{minipage}
\vfill
\begin{minipage}[h]{\textwidth}
  \centering
  \centerline{\includegraphics[scale =0.55]{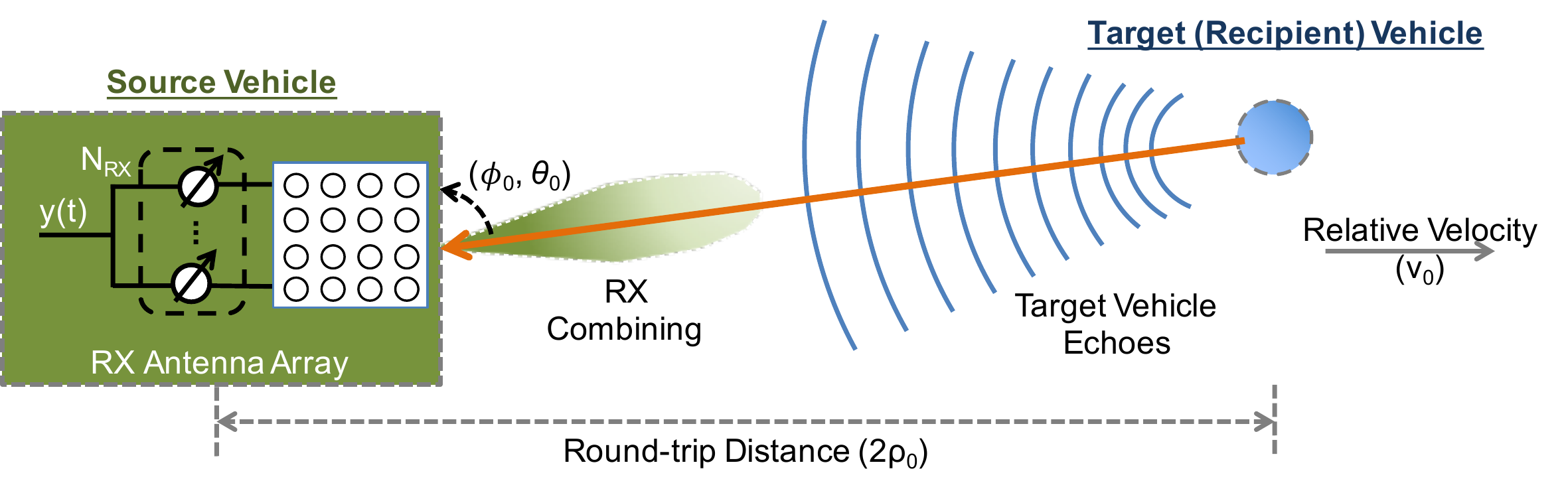}}
  {(b) Two-way radar channel for a single target vehicle, where the scattering centers of the recipient vehicle fall within a single radar resolution and are represented by a point target. Here, $y(t)$ denotes the continuous-time received radar signal.}\medskip
\end{minipage}
\caption{After the IEEE 802.11ad beam alignment, the TX and RX beams at the source vehicle are pointed towards the recipient vehicle.}
\label{fig_beamforming}
\vspace{-1.5em}
\end{figure}

IEEE 802.11ad supports multiple antenna communication with a single data stream. Spatial multiplexing as found in IEEE 802.11n/ac is not supported.  To develop a single data stream beamforming model, we incorporate the TX/RX analog beamforming vectors into the baseband model even though the actual beamforming may happen at an intermediate frequency (IF) or RF. We assume there is no blockage between the source and recipient vehicles.

We consider a coherent processing interval of $T$ seconds, where the location and velocity of a target vehicle, such as the recipient vehicle, is assumed to be constant. Therefore, the transmitted signal at the source vehicle during a CPI is
\begin{equation} \label{eq:tx_beam}
{\mathbf{x}_{\txm}}(t) = \wt x(t), \quad 0 \leq t \leq T
\end{equation}
where $\wt \in \mathbb{C}^{\Nt \times 1}$ is the TX frequency-flat analog beamforming vector at the source vehicle. The vector $\wt$ is time invariant in (\ref{eq:tx_beam}) because we assume the direction of the recipient vehicle is invariant within a CPI. 

\subsection{Channel and Target Models} \label{Sec:Channel}
The mmWave channel consists of contributions from a few scattering clusters \cite{RapMacSam:Wideband-Millimeter-Wave-Propagation:15} (such as reflections from the target vehicles and clutter) and from self- and inter-user interference. We use two-dimensional (2D) TX/RX antenna arrays at the source and the recipient vehicles because it will allow high-resolution beamforming in the azimuth and elevation directions and is used in mmWave communications (see, e.g., \cite{roh2014millimeter}) and automotive radars \cite{menzel2012antenna}. This will enable a large beamforming gain, mitigate inter-user interference, increase communication system capacity, and enhance resolution for radar sensing. In particular, we use uniform planar array (UPA) antennas with steering vector $\mathbf{a}(\phi,\theta)$ in azimuth angle $\phi$ and elevation angle $\theta$. UPAs are being considered in mmWave system design because of their high space efficiency acquired by placing antennas on a 2D grid (see, e.g.,\cite{SonChoLov:Common-Codebook-Millimeter:16}).


The TX and RX antenna arrays on the source vehicle are closely separated such that both arrays will see the same location parameters (e.g., azimuth/elevation angle and range) of a scatter. At the same time, the separation between the TX and the RX antenna arrays at the source vehicle along with the use of self-interference cancellation mechanism, TX/RX beamforming and an efficient circulator (e.g., \cite{estep2014magnetic}) will provide enough isolation and cancellation for full-duplex operation. Developing algorithms for full-duplex operation (e.g. self-interference mitigation in wireless-propagation-domain, analog-circuit-domain, and digital domain \cite{sabharwal2013band}) is a subject of future work. We also assume that the IEEE 802.11ad medium access control protocol will avoid the inter-user interference from other vehicles. 

During the mmWave joint communication-radar operation between the source and recipient vehicles, we assume that the 3-dB beamwidths of their TX and RX beams are narrow (as in \cite{wilocity15,roh2014millimeter}). We also assume that the beams are steered towards each other without any blockage. Although very narrow beams will lead to less clutter \cite[Ch. 7]{currie1987principles}, low interference \cite{saha201560}, and long range operation due to large beamforming gain, they can yield poor performance with vehicle mobility and blockage \cite{saha201560,va2015impact}. In \cite{va2015impact}, the trade-offs between the Doppler effect and the pointing error when choosing the beam width for mmWave vehicular communications have been studied, and it has been concluded that the beams must be pointy but not too narrow. Hence, we assume that the TX/RX beams are narrow enough to meet the link budget requirement of V2V communication and radar but are wide enough to illuminate all the scattering centers of a far target vehicle within their resolution (similar to the LRR beams defined in \cite{hasch2012millimeter}). Therefore, we represent the recipient vehicle as a single point target, as in \cite{rohling1996cfar, bazzi2012estimation}, and model the mmWave communication channel with a dominated LOS path corresponding to the recipient vehicle.

During a CPI, we assume that the recipient vehicle has an arbitrary range of $\rho_\um$ and azimuth/elevation direction pair of $(\phi_\um, \theta_\um)$ moving with a relative radial velocity $v_\um$ with respect to (w.r.t) the source vehicle, as shown in Fig.~\ref{fig_beamforming}. We also assume that the acceleration and the relative velocity of the recipient vehicle w.r.t the source vehicle is small enough to allow for constant velocity and quasi-stationary assumption for a CPI, that is, constant $v_\um$, $\rho_\um$, and $(\phi_\um,\theta_\um)$ \cite[Ch. 2]{richards2005fundamentals},\cite{rohling2001waveform}.

\subsubsection{Communication Channel Model}
To evaluate the trade-off between the communication data rate and radar estimation accuracy, we consider a single target scenario for simplicity. Assuming the recipient vehicle is the only dominant direct path scatter present in the radar channel, we model the one-way LOS dominant mmWave communication channel as a frequency-flat Rician channel \cite{VaShiBan:Millimeter-Wave-Vehicular:16}. This can be similarly extended to multi-target scenario by including frequency-selective communication channel model \cite{RapMacSam:Wideband-Millimeter-Wave-Propagation:15}. We assume that the channel is time-invariant during a single frame because the source and target vehicles are slow enough. We do not include band-limited filters in the channel model and instead include them in the TX/RX signal models. Additionally, the timing synchronization is considered in the received signal model (see Section~\ref{subSec:RxSig}).
The one-way LOS dominated small-scale communication channel corresponding to the $m^\thm$ frame in a CPI is represented as \cite{sayeed2010wireless}
\begin{equation}  \label{eq:channelComm}
\mathbf{H}_\comm[m] = \left( \sqrt{\frac{J_\comm}{J_\comm+1}} \mathbf{H}_{\mathrm{LOS}}[m] +   \sqrt{\frac{1}{J_\comm+1}} \mathbf{H}_\mathrm{w} [m] \right),
\end{equation}
where $J_\comm$ is the Rician factor, $\mathbf{H}_\comm[m] \in \mathbb{C}^{\Nr \times \Nt}$, and $\e{\vert \vert \mathbf{H}_\comm[m] \vert \vert ^2_{\mathrm{F}}} = \Nt \Nr$. The LOS channel matrix, $\mathbf{H}_\mathrm{LOS}[m]$, is expressed as
\begin{equation} \label{eq:channelLOS}
\mathbf{H}_\mathrm{LOS}[m] = \alpha_\um \me^{\jm 2\pi \nu_\um mK_m\Ts} \mathbf{a}_{\mrx}(\phi_{0},\theta_{0})\mathbf{a}^{\mathrm{*}}_{\txm}(\phi_{0},\theta_{0}),
\end{equation}
where $\alpha_\um$ is unit magnitude and fixed phase, $\mathbf{a}_{\txm}(\phi_{0},\theta_{0})$ denotes the TX steering vector at the source vehicle, $ \mathbf{a}_{\mrx}(\phi_{0},\theta_{0})$ is the RX steering vector at the recipient vehicle, $\nu_\um = {2 v_\um}/{\lambda}$ denotes the Doppler shift, and $\lambda$ represents the carrier wavelength. The DoA is same as the DoD in (\ref{eq:channelLOS}) because we consider the LOS channel. The elements of $\mathbf{H}_\mathrm{w}[m]$ are modeled by independent and identically-distributed (IID) complex Gaussian random variables with zero-mean and unit variance.
We assume that the source and recipient vehicles align their TX/RX beams toward each other using the IEEE 802.11ad beam training protocol. For the model in (\ref{eq:channelComm}), the TX beamforming vector, $\wt$, at the source vehicle and the RX beamforming vector, $\mathbf{f}_{\mrx,\comm}$, is chosen so that the beamforming gain is maximized \cite{zhou2012efficient}.
A particular TX/RX codebook is not specified in the IEEE 802.11ad standard. In numerical simulations, we adopt discrete Fourier transform (DFT)-based codebooks, which have been proposed for the practical implementations of mmWave WLAN systems \cite{zhou2012efficient}. Since the mmWave communication channel is LOS dominated, we assume that once the link has been established, the TX beam of the source vehicle and the RX beam of the recipient vehicle is assumed to be pointing towards $(\phi_\um,\theta_\um)$ with a small beam alignment error \cite{MarIwaOht:First-Eigenmode-Transmission:16}, as shown in Fig.~\ref{fig_beamforming}. 

The effective complex communication channel model after the TX/RX beamforming is expressed as
\begin{equation}
 h_\comm[m] = \sqrt{G_\comm}  {\mathbf{f}}^*_{\mathrm{\mrx,com}} \mathbf{H}_\comm[m] {\mathbf{f}}_\txm,
\end{equation}
where $G_\comm$ is the large-scale communication channel gain at the recipient vehicle. We use the close-in (CI) free space reference distance path loss model with CI free space reference distance of 1 m to model $G_\comm$, which leads to $G_\comm =  { \lambda^2 }/{(4 \pi)^2 \rho_0^\mathrm{PL}}$ \cite{RapMacSam:Wideband-Millimeter-Wave-Propagation:15}.
The exponent $\mathrm{PL}$ denotes the path loss (PL) exponent and is close to 2 for mmWave LOS outdoor urban \cite{RapMacSam:Wideband-Millimeter-Wave-Propagation:15} and rural channels \cite{ MacSunRap:Millimeter-Wave-Wireless:16}. The actual value of $\mathrm{PL}$, however, will depend on the specific vehicular scenario. In numerical simulations, we have studied the effect of $\mathrm{PL}$ on radar and communication performance. 
\subsubsection{Target and Clutter Model}

We model the mmWave radar channel for a single CPI using the doubly selective (time- and frequency-selective) model, which is used in automotive radar systems such as \cite{AskEkm:Tracking-With-a-High-Resolution:15}. The radar channel is assumed to be a sum of the contributions from a few $\Nl$ dominant direct path target echoes and multi-path spread-Doppler clutter. Each path corresponding to the $p^{\mathrm{th}}$ target echo is described by six physical parameters: its azimuth and elevation  angle of arrival (AoA)/angle of departure (AoD) pair $(\phi_{p},\theta_{p})$, round-trip delay, $\tau_{p}$, small-scale complex channel gain $\beta_{p}$, large-scale channel gain $G_{p}$, and Doppler shift $\nu_{p}$. The round-trip delay and Doppler shift corresponding to the $p^\thm$ target echo is related to its distance $\rho_p$ and relative velocity $v_p$ as $\tau_{p} = 2 \rho_p / c$, and $\nu_{p} = {2 v_p}/{\lambda}$, where $c$ is the speed of light. 

We represent the radar target model for the co-located TX/RX antenna arrays at the source vehicle as \cite{li2007mimo}
\begin{equation} \label{eq:channelMulTarget}
\mathbf{H}_\tm(t,f) = \sum_{p=0}^{\Nl-1} \sqrt{G_p} \beta_{p} \me^{\jm 2\pi \nu_{p}t} \me^{-\jm 2\pi \tau_{p}f} \mathbf{A}_{\mr}(\phi_{p},\theta_{p}),
\end{equation}
where $\mathbf{A}_{\mr}(\phi_{p},\theta_{p}) = \mathbf{a}^\cm_{\mrx}(\phi_{p},\theta_{p}) \mathbf{a}^{*}_\txm(\phi_{p},\theta_{p})$. The large-scale radar channel gain is assumed to follow free-space path-loss model with PL exponent of 2 (as used extensively in previous work, e.g., \cite{bazzi2012estimation}), i.e., $ G_p =  { \lambda^2 \sigma_{\mathrm{RCS},p}}/({64 \pi^3 \rho_p^4})$, where $\sigma_{\mathrm{RCS},p}$ is RCS corresponding to the $p^\thm$ target.  In (\ref{eq:channelMulTarget}), we only consider far target whose $\rho_p$ is large compared to the distance change during the CPI, i.e., $\rho_p \gg v_p/T$. Hence, we assume constant $\beta_p$ \cite{rohling2001waveform}. We have not included the band-limiting filters in (\ref{eq:channelMulTarget}) because they have been taken care in the TX signal model in (\ref{eq:TXCont}). In the case of $\Nl =1$, (\ref{eq:channelMulTarget}) will represent a single-target model and for $\Nl >1$, (\ref{eq:channelMulTarget}) will represent a multi-target model. Specifically, we focus on the physical parameters of the $0^\thm$ path representing the direct path between the source and recipient vehicles for both single- and multi-target models. 

Therefore, the two-way radar channel with the multi-path spread-Doppler clutter matrix, $\mathbf{H}_\cm (t,f)$, is represented as
\begin{equation} \label{eq:channelRadar}
\mathbf{H}_\mr(t,f)  = \mathbf{H}_\tm (t,f) + \mathbf{H}_\cm (t,f),
\end{equation}
where $\mathbf{H}_\cm (t,f)$ is assumed to be an IID complex Gaussian distributed random process because of constant TX/RX beamforming vectors (non-scanning mode) during a CPI \cite[Ch. 2 and 7]{currie1987principles}. This assumption is also used in automotive radar algorithms, such as \cite{rohling1996cfar,LimJeoLee:Rejection-of-road-clutter:11}. If the clutter component is not Gaussian, space-time adaptive processing can be used as a preprocessing step to filter out the clutter component \cite{Kle:Applications-of-space-time-adaptive:04}. The distribution of the elements of $\mathbf{H}_\cm (t,f)$ impacts the choice of algorithm and performance bounds for radar detection and parameter estimation. 

We assume that at both the source and recipient vehicles, the same IEEE 802.11ad-based beamforming codebook is used. The AoA pair at the radar receiver mounted on the source vehicle and the AoA pair at the communication receiver mounted on the recipient vehicle is the same for the direct path between the source and the recipient vehicles, as shown in Fig.~\ref{fig_beamforming}. Therefore, it is reasonable to assume that the radar RX beamforming vector, $\mathbf{f}_{\mrx,\mr}$, at the source vehicle with the target vehicle model in (\ref{eq:channelMulTarget}) and the communication RX beamforming vector, $\mathbf{f}_{\mrx,\mr}$, at the recipient vehicle with the LOS channel matrix in (\ref{eq:channelLOS}) satisfies 
${\mathbf{f}}_{\mrx,\mr} = {\mathbf{f}}^\cm_{\mrx,\comm}$.
This assumption will enable us to compare the received power between the radar receiver at the source vehicle and the communication receiver at the recipient vehicle.

After the TX/RX beamforming, the effective target model is $h_\tm (t,f) = {\mathbf{f}}^*_{\mathrm{\mrx,\mr}} \mathbf{H}_\tm(t,f) {\mathbf{f}}_\txm$ and the effective clutter model is $h_\cm (t,f) = {\mathbf{f}}^*_{\mathrm{\mrx,\mr}} \mathbf{H}_\cm(t,f) {\mathbf{f}}_\txm$. In particular, the effective multi-target model is non-linearly dependent on the physical parameters, making it difficult to analyze and estimate the multi-target parameters. The target delays and Doppler shifts during a CPI, however, can be well approximated using the linear counterpart of $h_\tm (t,f)$, known as the 2D delay-Doppler map, $H_\mapm[\ell,d]$ \cite[Ch. 7]{richards2005fundamentals}. The delay-Doppler map partitions the $\Nl$ paths into a 2D resolution cell of size $\Delta \tau \times \Delta \nu$, where $\Delta \tau = 1/W$ and $\Delta \nu = 1/T$. Assuming $\tau_{\mathrm{max}}$ is the maximum delay spread and $\nu_{\mathrm{max}}$ is the maximum Doppler spread during the CPI, the maximum number of delay resolution bins is $L= \lceil W \tau_{\mathrm{max}} \rceil + 1$ and the maximum number of resolvable (one-sided) Doppler shifts is $D=\lceil T \nu_{\mathrm{max}}/2 \rceil$. Therefore, instead of representing the channel using continuous delay and Doppler, the delay-Doppler map is represented by uniform spaced delays $\tau_{\ell}=\ell/W$, and Doppler shifts $\nu_{d}=d/T$. The virtual delay-Doppler map representation of $h_\tm (t,f)$ uniformly sampled in delay and Doppler dimensions commensurate with the resolution in their respective dimensions is given by \cite{sayeed2010wireless}
\begin{equation} \label{eq:map}
h_\tm(t,f) \approx \sum_{\ell=0}^{L-1} \sum_{d=-D}^{D}  H_\mapm[\ell,d] \me^{- \jm 2\pi \frac{\ell}{W}f} \me^{\jm 2\pi \frac{d}{T}t} .
\end{equation} 
We apply classical low-complexity pulse-Doppler algorithms on the 2D delay-Doppler map obtained during a CPI to estimate target delays and Doppler shifts \cite[Ch. 7]{richards2005fundamentals}, as explained later in Section \ref{Sec:ProRx}. 
\subsection{Received Signal Model} \label{subSec:RxSig}
We consider a coherent processing interval of duration $T$ seconds containing $M$ frames. For simplicity, we assume each frame in the CPI consists of $K$ samples meaning that the data payload is of the same size. This will allow us to leverage the range and velocity estimation algorithms of a classic pulse-Doppler radar, which has a constant pulse repetition frequency, for developing multi-frame IEEE 802.11ad-based radar processing techniques. This assumption also simplifies the received signal model. The assumption holds true when the communication system uses maximum frame length for a given channel delay and Doppler spread during high data transmission load scenario. The radar processing techniques, however, can be extended for different frame lengths within a CPI. 


\emph{Communication Received Signal Model:} After matched filtering with $g_\mrx(t)$, time/frequency synchronization, and symbol rate sampling, the discrete-time received communication signal at the recipient vehicle corresponding to the $k^\thm$ symbol in the $m^\thm$ frame during a CPI is represented as
 \begin{equation} \label{eq:finalSigComm}
y_\comm[k,m] = \sqrt{\Es } h_\comm[m] s[k+mK] +z_{\comm}[k,m],
\end{equation} 
where $z_{\comm}[k,m]$ is the AWGN noise, which is distributed as $\mathcal{N_C}(0,\sigma_{\nm}^2)$. The SNR of the received communication signal at the recipient vehicle is defined as $\snr = {\Es \vert h_\comm[m] \vert ^2}/{\sigma_{\nm}^2}$. 

\emph{Radar Received Signal Model:} We apply the stop-and-hop assumption to model the round-trip delay and phase modulation in a time-varying echo signal \cite[Ch. 2]{richards2005fundamentals}. Under this assumption, the echo is received with a time delay corresponding to the range at the beginning of the pulse transmission but with a phase modulation related to the time variation in range. Then, the received radar signal after matched filtering with $g_\mrx(t)$ for a single target model during a CPI will result in
 \begin{equation} \label{eq:finalSigCont}
y(t) =  \sqrt{\Es} \hum  x_\psm(t-\tau_\um) \me^{\jm 2\pi \nu_\um t}  + z_{\cm}(t)+z_{\nm}(t),
\end{equation} 
where $h_\um = \sqrt{G_\um} \beta_\um {\mathbf{f}}^*_{\mathrm{\mrx,\mr}}  \mathbf{A}_{\mr}(\phi_{\um},\theta_{\um}) {\mathbf{f}}_\txm $, $x_\psm(t-\tau_\um) = \sum_{n =-\infty}^{\infty} s[n]g(t - n \Ts - \tau_\um)$, $g(t) = g_\txm(t) \ast g_\mrx(t)$, and ${z}_{\nm}(t)$ is the complex gain additive white Gaussian noise (AWGN) with power $\sigma^2_{\mathrm{n}}$. The clutter term $z_{\cm}(t)$ is assumed to be distributed as $\mathcal{N_C}(0,\sigma^2_{\mathrm{c}})$ because ${h}_\cm(t,f)$ is an IID complex Gaussian random process. 

The discrete-time representation of the received training sequence with length $K_\trm$ corresponding to the $k^{\mathrm{th}}$ symbol in the $m^{\mathrm{th}}$ frame during a CPI is
\begin{equation} \label{eq:finalSig}
y[k,m] =  \sqrt{\Es} \hum \me^{\jm 2\pi \nu_0 (k+m K) \Ts} x_{\psm}(k\Ts -\tau_\um) +z_{\cm\nm}[k,m],
\end{equation}
where $y[k,m] = y((k+mK)\Tc)$, and $x_\psm(k\Ts -\tau_\um) = x_\psm((k+mK)\Ts -\tau_\um)$, i.e., the transmitted training symbols are the same across all the $M$ frames. The clutter-plus-noise term $z_{\cm\nm}[k,m] =z_{\cm}((k+mK)\Tc)+ z_{\nm}((k+mK)\Tc)$ is assumed to be distributed as $\mathcal{N_C}(0,\sigma_{\cm\nm}^2)$, where $\sigma_{\cm\nm}^2 = \sigma_{\cm}^2W + \sigma_{\nm}^2W$. The SCNR of the received radar signal at the source vehicle can, therefore, be defined as $\scnr = \Es \vert \hum  \vert^2/\sigma_{\cm\nm}^2$. The received signal model for a single target can be extended to multi-target vehicular scenario by adding more terms corresponding to multiple targets based on $H_\mapm[\ell,d]$, as explained in \cite[Ch. 2]{richards2005fundamentals}.
\section{Proposed Receiver Processing Techniques For Enabling Radar Functions} \label{Sec:ProRx}
\begin{figure}[!t]
\centering
\includegraphics[scale =0.5]{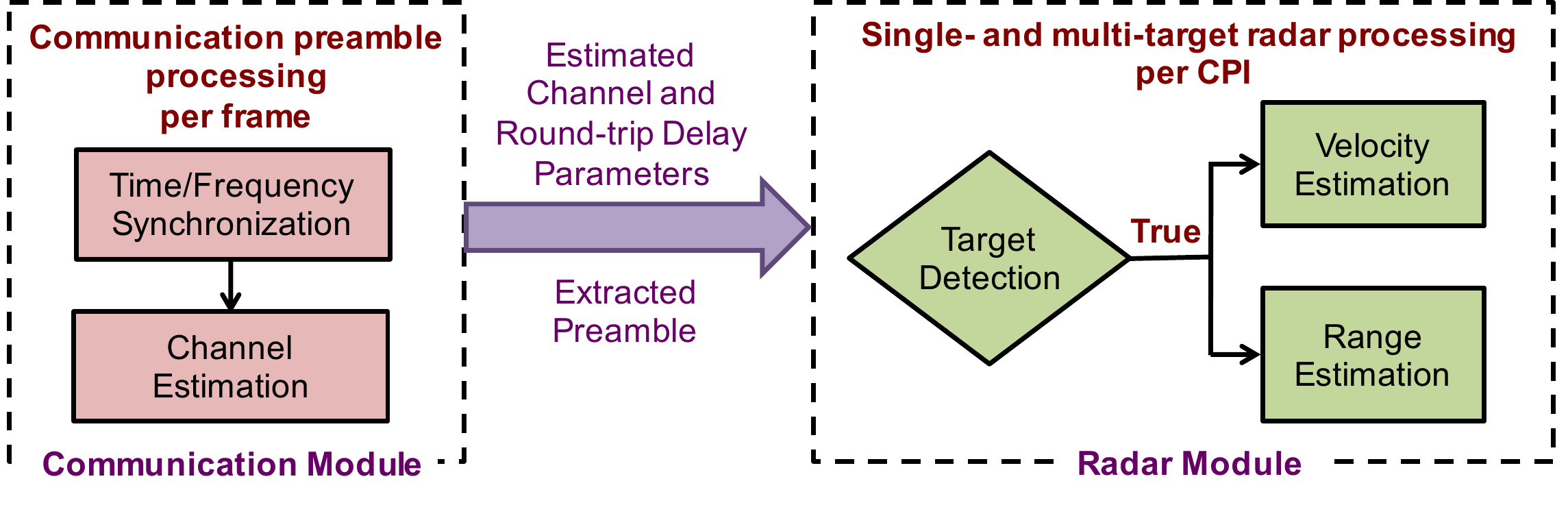}
 \caption{The processing techniques for target detection and estimation of range and velocity using an IEEE 802.11ad-based joint communication-radar system. The processing algorithms leverage the STF and the CEF of multiple frames in a CPI.}
\label{fig_flowchart}
 \vspace{-0.5cm}
\end{figure}

We propose an IEEE 802.11ad-based radar receiver at the source vehicle that consists of a communication module and a radar module, as shown in Fig. \ref{fig_flowchart}. We consider three primary types of processing in the radar module: 1) \emph{vehicle detection} using a constant false alarm rate algorithm;  2) \emph{range estimation} using time synchronization techniques; and 3) \emph{velocity estimation} using frequency synchronization techniques. The algorithms used in the radar processing module are developed by extending the communication processing techniques over a single frame to multiple frames in a CPI. This approach will enable the realization of a joint communication-radar system using a conventional low-cost IEEE 802.11ad system with minimal receiver modifications.
\subsection{Communication Preamble Processing per Frame} \label{Sec:Comm}
In the communication module, training sequences in the preamble of a single frame are used for time/frequency synchronizations and channel estimation \cite{liuall}. This is achieved in several steps: 1) \emph{coarse time synchronization} based on preamble detection techniques using the STF; 2) \emph{frequency offset estimation} using the STF; 3) \emph{fine time synchronization} using the CEF symbol boundary detection and the STF/CEF peak detection techniques;  and 4) \emph{channel estimation} using the CEF. The accuracy of Doppler shift estimation using a single frame in the communication module is inaccurate at low SNR due to small Doppler shift and less integration time, as explained later in Remark 2. Therefore, we estimate the Doppler shift in the radar module using multiple frames.

For simplicity, we describe the processing techniques of the communication module for a single target vehicular scenario, which can be extended to a multi-target vehicular situation. The first step of the training sequence processing is the timing synchronization. Since the round-trip delay of a target vehicle, $\tau_\um$, is a continuous variable, we can represent it as $\tau_\um = \ell_0 \Ts + \taud$, where $\ell_0$ is an integer and $\taud$ is a fractional symbol delay. 
Then,  (\ref{eq:finalSig}) can be represented for $0<k<K_\trm$ as
\begin{equation} \label{eq:finalSig1} 
y[k,m] = \sqrt{\Es} \hum s[k-{\ell_\um}] g (\tau_\um)\me^{\jm 2\pi \nu_\um (k+mK) \Ts}  \\ + z_\isim[k,m]  + z_{\cm\nm}[k,m] , 
\end{equation} 
where, $ z_\isim[k,m] = \sqrt{\Es}\hum \sum \limits_{n \neq k+mK}s[n]g(((k+mK)-n)\Tc -\tau_\um) \me^{\jm 2\pi \nu_\um (k+mK) \Ts}$ is the intersymbol interference (ISI).

We use the energy-based symbol synchronization algorithm to estimate the fractional symbol delay, $\hat{\taud}[m]$, and then apply a fractional symbol delay correction to mitigate its effect \cite{liuall,preyss2015digital}. Since the assumed TX and the RX pulse shaping RRC filters lead to an equivalent filter satisfying the Nyquist condition, we consider $g(n\Tc) = \delta[n]$. Therefore, the received signal corresponding to the $m^{\mathrm{th}}$ frame is 
\begin{equation} \label{eq:finalSigProc}
{y}_m[k] =  \sqrt{\Es} \hum \me^{\jm 2\pi \nu_0 (k+m K) \Ts} s[k-{\ell_\um}] +{z}_m[k],
\end{equation}
where $\mathbf{y}_m \in \mathbb{C}^{K_\trm \times 1}$ represents the vector of the received training symbols with $y_m[k] = y[m,k]$, and $ \mathbf{{z}}_{m} \in \mathbb{C}^{K_\trm \times 1}$ represents the residual ISI-plus-clutter-plus-noise vector (i.e., $z_m[k]$ is the sum of $z_{\cm\nm}[k,m]$ and the residual ISI after fractional symbol delay correction).

After symbol synchronization, we detect the IEEE 802.11ad frame using the normalized auto-correlation of the STF, which consists of 16 repeated $\Ga$. The $\ell^{\mathrm{th}}$ normalized auto-correlation corresponding to the $m^\thm$ frame is given by
\begin{equation} \label{Rjoint}
\Rj[\ell,m] = \frac{ \sum_{n=0}^{\NT-1} y_m[\ell-n] {y_m^*}[\ell-n-\ND]}{ \sqrt{\sum_{n=0}^{\NT-1}  \vert {y}_m{[\ell-n]}  \vert ^2}    \sqrt{\sum_{n=0}^{\NT-1} \vert {y_m^*}{[\ell-n-\ND]}  \vert ^2 }} ,
\end{equation}
where $\NT=128$ is the length of the training sequence and $\ND=128$ is the distance between the consecutive training sequences chosen for correlation. The frame start is detected when $\vert \Rj[\ell,m] \vert > \chi_{\mathrm{STF}}$ for 128 times, where $\chi_{\mathrm{STF}}$ is a pre-defined threshold and is $< 1$ \cite{liuall}. The frame start detection technique, therefore, uses around $128 \times 2$ to $128 \times 3$ samples in the STF field to confirm the detection. The coarse range estimate of the target vehicle by applying the preamble start detection technique to the $m^{\mathrm{th}}$ frame is given by  
\begin{equation} \label{eq:frameStart}
\hat{\ell}_\mathrm{01}[m] = \mathrm{inf} \left\lbrace {\ell} \mid \vert \Rj[\ell,m] \vert \geq \chi_{\mathrm{STF}} \right\rbrace.
\end{equation}


The carrier frequency offset (CFO) can be estimated by using the auto-correlation based algorithm and residual CFO estimation techniques proposed in \cite{liuall}. For simplicity, we assume that these algorithms achieve perfect carrier frequency offset compensation.

The fine range estimate of the time-delay can be obtained either by using an amplitude-based method or a phase-based method. The amplitude-based method estimates the fine time-delay using the cross-correlation, $R_{2}[\ell,m]$, between multiple $\Ga$ in the STF sequence, which is expressed as
\begin{equation}
 R_{2}[\ell,m] = \sum_{i=0}^{\Np-1} \sum_{n=0}^{\NT -1} a_{\mathrm{128}}[n] y_m[\ell+n+i\NT],
 \end{equation}
where $\NT= 128$, and $\Np = 16$ is the total number of repetitions of $\Ga$ in the STF. The fine-time delay, ${\hat{\ell}}_{02}[m]$, is estimated by detecting the peak of $R_{2}[\ell,m]$ and is given by
\begin{equation}
{\hat{\ell}}_{02}[m] = \argmaxA_{\ell: \ell \in \mathbb{Z}}  \vert R_2[\ell,m]  \vert^2,
 \end{equation}
 where $\mathbb{Z}$ is the set of integers. 
The amplitude-based fine timing synchronization can also be similarly performed by applying the peak detection technique on the CEF instead of the STF.  Both the peak detection methods perform well even when SCNR $\scnr$ is low.

The timing synchronization at the $m^{\mathrm{th}}$ frame can also be fine tuned by performing phased-based CEF symbol boundary detection \cite{liuall}.
This method, however, does not perform well in the presence of Doppler shift at low SNR of the received communication signal. 

After the fine time synchronization, we extract the received CEF signal to estimate the channel using the 512 sample Golay complementary pair. The channel estimate, $\hat{h}_m[\ell]$, is acquired after removing the cyclic prefix from the correlation values, $\hat{\gamma}(\mathbf{y}_m ,\ell)$, between the received CEF and $[\Gu \; \Gv]$ \cite{liu2013digital}. Therefore, $\hat{\gamma}(\mathbf{y}_m ,\ell)$ and $\hat{h}_m[\ell]$ are expressed as
\begin{equation} 
\hat{\gamma}(\mathbf{y}_m ,\ell) =\frac{1}{2\NT}  \Bigg ( \sum^{\NT-1}_{n=0} {y}_m[n+\ell] {a}_{512}^*[n]  +  \sum^{\NT-1}_{n=0} {y}_m[n+\ell+\NT] {b}_{512}^*[n]  \Bigg )  \end{equation}  
 and
\begin{equation} \label{eq:channelEstM}
\hat{h}_m[\ell]= \hat{\gamma}(\mathbf{y}_m ,\ell+N_{\mathrm{CP}})  \quad \ell = 0,\cdots,\NT-1 ,
\end{equation}
where $P=512$ and the length of the cyclic prefix $N_{\mathrm{CP}} = 128$. 

We assume the channel is time invariant during the CEF because the source and recipient vehicles are slow enough. Based on the channel model in (\ref{eq:channelRadar}) with a single target, the channel estimate in (\ref{eq:channelEstM}), which leverages the perfect auto-correlation property of Golay complementary pair, can be decomposed as \cite{liu2013digital}
\begin{equation} \label{eq:detectChannel}
\hat{h}_m[\ell] = \begin{cases}
\sqrt{\Es} \hum \me^{- \jm 2\pi \nu_\um m K\Ts} +  {{\tilde{z}}}_m[\ell] & \ell=\ell_{\cefm} \\
{\tilde{z}}_m[\ell]  & \text{otherwise}
\end{cases}
\end{equation} 
where $ {\tilde{z}}_m[\ell]=  \hat{\gamma}( {\mathbf{z}}_m ,\ell+N_{\mathrm{CP}})$ and $\ell_{\cefm} = 256$. Note that ${z}_m[k]$ represents the residual ISI-plus-clutter-plus-noise term as defined in (\ref{eq:finalSigProc}).

\subsection{Single Target Radar Processing per CPI}
The radar module leverages the training sequence processing in the communication module to detect and estimate the range and velocity of the target vehicle, which is the recipient vehicle, for a single target model.

\subsubsection{Target Detection}
The target vehicle can be detected by applying a constant false alarm rate (CFAR) detection technique either on the channel estimate in (\ref{eq:detectChannel}) or on the energy of the cross-correlation between the received and transmitted preambles, $E_\mathrm{pream}$ (as used in classic radar detection \cite{Roh:Some-radar-topics::06}).
In the CFAR technique, the decision is based on a simple thresholding function
\begin{equation} \label{eq:qD}
  \varphi(E) = 
\begin{cases}
    0 & \text{if } E  < \chi_\mathrm{D}  \\
   1 & \text{if } E  > \chi_\mathrm{D}   .
\end{cases}
\end{equation}
where, $E = \vert \hat{h}_m[\ell_\cefm] \vert^2$ for CEF-based estimation and $E = E_\mathrm{pream}$ for preamble-based estimation.
For a constant false alarm probability of $P_{\mathrm{FA}}$, the detection threshold becomes 
$\chi_{\mathrm{D}} = -\sigma^2_{\cm\nm} \mathrm{ln} P_{\mathrm{FA}}$ \cite[Ch. 6]{richards2005fundamentals},
where $\sigma^2_{\cm\nm}$ is the variance of the zero-mean complex Gaussian clutter-plus-noise term ${\tilde{z}}_m[\ell]$. We assume that the value of $\sigma^2_{\cm\nm}$ is known because it can be calculated using the typical mmWave WLAN noise variance estimation technique \cite{bo2015compressed}. The target detection using the entire preamble will achieve higher $P_\mathrm{D}$ at a given $P_\mathrm{FA}$ at the expense of higher sidelobes, which is especially unfavorable for multi-target scenario. Therefore, we use the preamble for detecting a single target and use the CEF for multi-target detection provided the SNR is high enough.

\subsubsection{Range Estimation} \label{subSec:RangeEst}
Once the target vehicle is detected at the source vehicle, the target range is calculated from its corresponding round-trip delay estimate. The range estimation algorithms are applied on the STF and the CEF and can be categorized into coarse and fine range estimation techniques. The coarse range estimation using a frame start detection algorithm estimates $\hat{\ell}_{01}[m]$ with an error of less than $128 \times 3$ samples. Fine range estimation based on the symbol boundary detection or the STF/CEF peak detection and symbol synchronization techniques result in the delay estimate of ${\hat{\ell}}_{02}[m] +\hat{\taud}[m]$ or ${\hat{\ell}}_{03}[m]+\hat{\taud}[m]$ with an error of less than 1 sample \cite{liuall}, which meets the LRR specification of $0.1$ m range accuracy\cite{hasch2012millimeter}.

\begin{remark}
The CRLB bound of the range estimation using the IEEE 802.11ad preamble can be expressed following the approach in \cite[Ch. 7]{richards2005fundamentals} as 
\begin{equation} \label{eq:crlb_range}
\sigma^2_{\hat{\rho}} =  \frac{c^2}{8 \eta^2 W^2 P \scnr},
\end{equation}
where $\eta$ depends on the power spectral density shape of $x(t)$ over the preamble duration. We assume a flat spectral shape of the preamble, which will allow better channel equalization of the communication system (e.g., Zadoff-Chu sequences used in LTE) and better radar parameter estimation of the target vehicle (e.g., linear frequency modulated chirp used in automotive radar). Due to the assumption of flat spectral shape, $\eta^2 = (2\pi)^2/12$ \cite[Ch. 7]{richards2005fundamentals}. The integration gain $P$ is equal to the number of preamble symbols used for range estimation, i.e., $16 \times 128$ for the fine range estimation using the STF and $8 \times 128$ for the fine range estimation using the CEF. The range estimation CRLB, as can be seen from (\ref{eq:crlb_range}), decreases with the integrated SCNR, which is defined as $P \scnr$.
\end{remark}

For $\scnr > 0$ dB and bandwidth $W > 1.76$ GHz (the exact value of $W$ will depend on chosen pulse shaping filter), it can be calculated from (\ref{eq:crlb_range}) that it is possible to achieve less than 1 mm accuracy using a single IEEE 802.11ad frame. In particular, for the RRC pulse shaping filter that we use in the numerical simulations, the range accuracy is 0.8 mm.

\subsubsection{Velocity Estimation} \label{Sec:Vel}
The relative velocity of the target vehicle is calculated at the source vehicle by estimating the Doppler shift of the corresponding target echo. We use the least squares (LS)-based frequency-offset estimation method over single/multiple frames to estimate the Doppler shift corresponding to the target vehicle. For this purpose, we choose $\mathbf{p} \in \mathbb{C}^{\Pt M \times 1}$ to be a vector of $M$ frames across $P$ delay bins, i.e., \\
\begin{equation} \label{eq:pVelVector}
\mathbf{p} = \Big[ [ y_0[k_0], \cdots, y_0[k_{P-1}] \cdots  y_{M-1}[k_0], \cdots, y_{M-1}[k_{P-1}]  \Big]
\end{equation}
where $\left\lbrace {k_i} \mid 0 \leq i \leq \Pt-1 \right \rbrace$ is an index set to the location of the training sequences in each frame.

The Doppler frequency estimation based on the Moose algorithm \cite{moose1994technique} or the CFO estimation algorithm used in IEEE 802.11ad \cite{liuall}, when applied on a single frame does not achieve the desired velocity accuracy of 0.1 m/s due to the small integration time, $T_{\mathrm{int}} = P \Ts$, for velocity estimation, as shown in \cite{preeti2015}. Therefore, to achieve desired velocity accuracy, we propose a multi-frame Moose-based algorithm for the Doppler frequency estimation problem as
\begin{equation} \label{eq:freqFinal}
\hat{\nu}_{\um} = \frac{\angle{\left( \sum_{i=0}^{M-1} \sum_{n=0}^{\NT-1} p[n+\ND +iP] {p^*}[n+ i\NT]\right)}}{2 \pi T_{\mathrm{D}}},
\end{equation}
where $\ND$ is the distance between two training sequences chosen for correlation and $T_{\mathrm{D}}$ is the time interval between these two training sequences, i.e., $\ND \Ts$. In the case of multi-frame velocity estimation, we choose $\ND = K$. In the case of single-frame velocity estimation, we choose $\ND = 4 \times 128$. Choosing larger $\ND$ improves the estimate, whereas it reduces the range of offsets that can be corrected, as explained later in Remark 2. The accuracy of frequency-offset estimation will improve when we use multiple frames (similar to pulse-Doppler radar) as compared to a single frame (traditionally used in frequency synchronization algorithms of a standard WLAN receiver) because of larger integration time, $ T_{\mathrm{int}} = M\TD$. The length of an IEEE 802.11ad frame SCPHY frame is variable from 0.002 ms to 1.2 ms \cite{ieee2012wireless}, whereas CPI of $T=10$ ms \cite{rohling2001waveform} and update rate of 10 Hz \cite{saponara2014highly} is used in automotive radar. Therefore, we can use multiple frames for radar processing, where the number of frames is $T/(K\Ts)$. 


\begin{remark} The theoretical performances of the proposed velocity estimation for a single target vehicle with velocity $v$ in a flat fading channel are summarized as follows.
\begin{enumerate}[(a)]

\item The CRLB for the velocity estimation using the STF of a single frame, as described in Appendix A, is 
\begin{equation} \label{eq:vel-crlb1}
\sigma^2_{\hat{v}} = \frac{6 \lambda^2}{(4\pi)^2 \NT^3\Ts^2 \scnr}.
\end{equation}
The CRLB expresses a lower bound on the variance of velocity estimators using the STF of a single frame. If $\sigma^2_{\hat{v}}$ is above the LRR's desired MSE for velocity estimation, then it indicates that the requirement for LRR velocity accuracy cannot be met by any unbiased estimator. It can be inferred from (\ref{eq:vel-crlb1}) that the velocity MSE decreases rapidly with an increase in $P$ and $\scnr$. The value of $P$, however, is constant in an SC PHY frame and is equal to 128 $\times$ 16, which implies that CRLB is mainly affected by the change in $\scnr$.

\item The CRLB for velocity estimation using preamble across multiple frames for large $M$, as derived in Appendix A, is
\begin{equation} \label{eq:vel-crlb2}
\sigma^2_{\hat{v}} \approx  \frac{6 \lambda^2}{(4\pi)^2(M P^3 + M^3 P K^2)\Ts^2 \scnr}.
\end{equation}
Similar to (\ref{eq:vel-crlb1}), (\ref{eq:vel-crlb2}) also suggests that estimated velocity accuracy enhances with the increase in number of preambles, i.e., $MP$, and $\scnr$. Unlike (\ref{eq:vel-crlb1}), however, (\ref{eq:vel-crlb2}) adds the flexibility of increasing the total number of total preamble symbols by choosing higher values of $M$, which improves the accuracy of the velocity estimation. 

The extra flexibility in varying $M$ due to the use of multiple frames enables a system trade-off between target velocity estimation accuracy and communication data rate for the number of frames within a CPI. The velocity estimation CRLB decreases with an increase in the total training sequence duration and the numbers of frames within a fixed size CPI, as can be seen from (\ref{eq:vel-crlb2}). 
The number of communication data symbols and consequently data rate, however, decreases with an increase in the training sequence duration. 

\item Due to the periodicity of the exponential function, the estimate of the Doppler shift calculated in (\ref{eq:freqFinal}) will only be accurate for 
\begin{equation} \label{eq:velAmb}
\vert \hat{\nu}_\um \vert \leq \frac{1}{2\ND \Ts}.
\end{equation}
In (\ref{eq:freqFinal}), we can use different periodicity of the preamble by choosing different training sequences. Comparing (\ref{eq:vel-crlb1}), (\ref{eq:vel-crlb2}) and (\ref{eq:velAmb}), however, we infer that there is a trade-off between accuracy and span of the unambiguous velocity estimation. The multi-frame Doppler estimation with $\ND > 128 \times 26$ has a higher accuracy as compared to the single-frame Doppler estimation with $\ND = 512$, whereas it has a comparatively reduced range of Doppler offsets that can be corrected due to larger $\ND$. 
\end{enumerate}
\end{remark}
\subsection{Multi-target Radar Processing per CPI} \label{Sec:Mul}
In the case of a multi-target model, we use classic pulse-Doppler-based radar processing technique that leverages the channel estimate derived in (\ref{eq:channelEstM}) to estimate the range and velocity of multiple targets. First, we obtain an estimate of the delay-Doppler map, $\hat{H}_\mapm[\ell,d]$, where the $\ell^\thm$ row of the 2D map, $\hat{H}_\mapm[\ell,:]$, is the $M$-point DFT of the zero-padded channel estimate vector $\left[ \hat{h}_0[\ell], \hat{h}_1[\ell], \cdots, \hat{h}_{M-1}[\ell] \right]$. Then, we use a thresholding method similar to (\ref{eq:qD}) to detect multiple targets from the delay-Doppler map \cite{Roh:Some-radar-topics::06}. The range of the $p^\thm$ detected target is estimated from the location of its corresponding delay bin, $\hat{\ell}_p$, in $\hat{H}_\mapm[\ell,d]$. Similarly, we determine each target's velocity from its corresponding Doppler bin $\hat{d}_p$. We use the channel estimates, which are derived from the CEF in $M$ frames, for multi-target radar processing because they exploit the desirable perfect auto-correlation property of the Golay complementary pair. 

\begin{remark}
The range resolution for the multi-target radar processing is
$\Delta \rho = c/(2 W)$ \cite[Ch. 10]{skolnik2008radar}.

The velocity resolution for multi-target model using conventional Fourier processing is $\Delta v =  {\lambda}/({2 T_{\mathrm{int}}})$, where $T_{\mathrm{int}} = M\TD$ with $\TD = K\Ts$ \cite[Ch. 10]{skolnik2008radar}. This implies that as the number of frames increases, the resolution of the velocity estimation increases.
\end{remark}
For $W > 1.76$ GHz and $T > 4.2$ ms, it can be calculated from the theoretical bounds in Remark 3 that it is possible to achieve less than 8.52 cm range resolution and less than 0.6 m/s of velocity resolution using the IEEE 802.11ad preamble, which is better than the required LRR resolution specifications in \cite{hasch2012millimeter} and typical CPI duration used in automotive radars (see, e.g., 10 ms CPI in \cite{rohling2001waveform}). In particular, for the RRC pulse shaping filter that we use in the numerical simulations, $\Delta \rho =$ 7 cm. In numerical results, we show that our proposed joint system and algorithms achieve these bounds.

\section{Numerical Results}
In this section, we perform Monte-Carlo simulations with 10,000 trials to evaluate the proposed radar techniques using IEEE 802.11ad against the required system specifications for LRR in a typical automotive radar setting \cite{hasch2012millimeter}. We assume the vehicle radar cross section is 10 dBsm \cite{bazzi2012estimation}. The multiple-antenna system is assumed to be a UPA with 8 horizontal and 2 vertical elements, as used in the Qualcomm IEEE 802.11ad chipsets \cite{Zhu:2014:DOP:2639108.2639121}. The 3-dB horizontal beamwidth of the UPA is $13^{\circ}$, and the 3-dB vertical beamwidth of the UPA is $60^{\circ}$.

\subsection{Single Target Automotive Scenario}
We simulate the received radar signal for a single target vehicular scenario, where the recipient vehicle is assumed to be the only target vehicle, as discussed in Section \ref{sec:System}. We choose the distance and the relative speed between the target and source vehicles as 50 m and 20 m/s, which falls in the typical span of LRR range and velocity specifications \cite{hasch2012millimeter}. We chose fixed location parameters because the performance bounds of the radar detection and estimation parameters do not vary with range and relative velocity of the target vehicle, as discussed in Remark-1 and Remark-2 in Section \ref{Sec:ProRx}. 

\begin{figure}[!t]
\centering
\includegraphics[scale = 0.65]{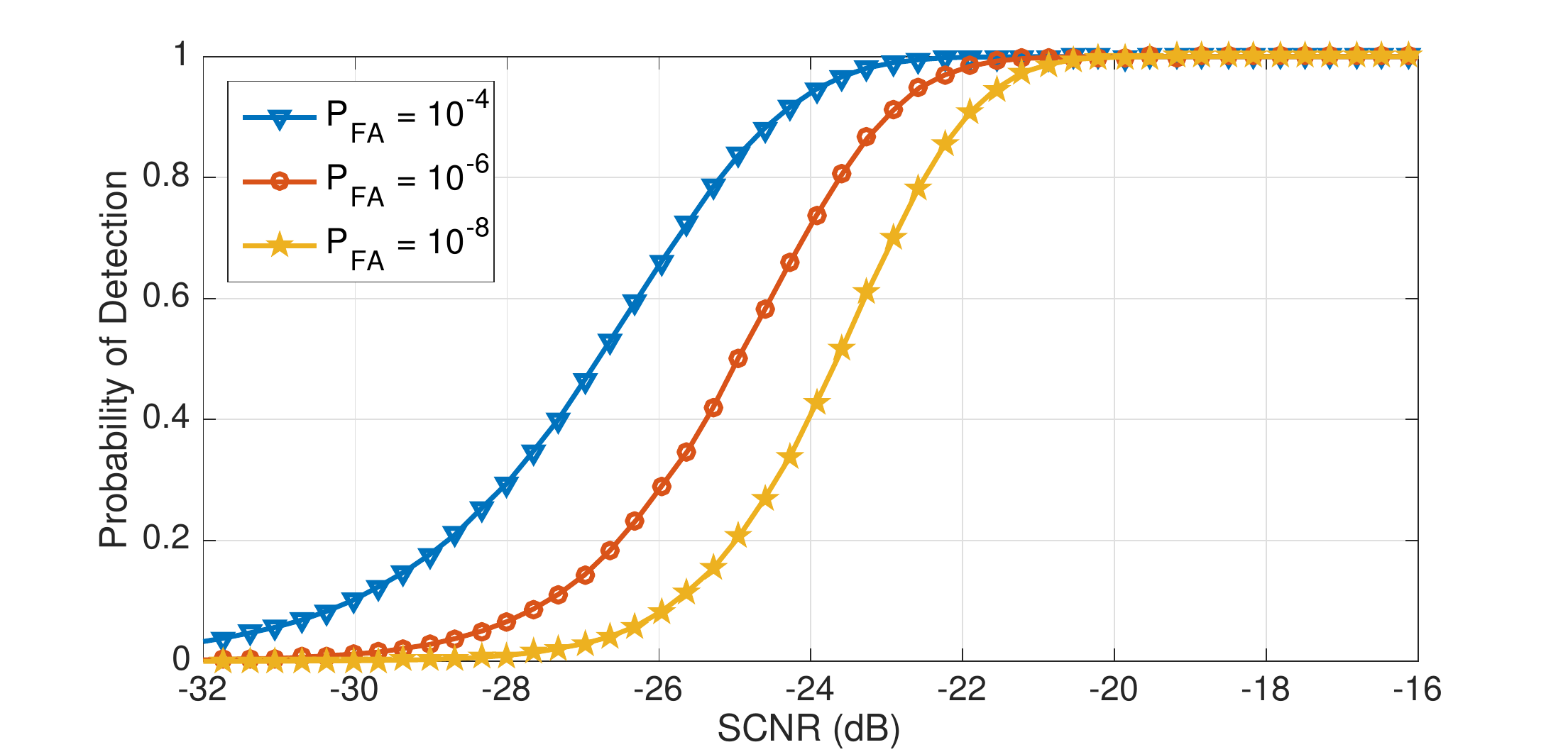}
 \caption{Probability of detection using different constant false alarm detection rates.}
\label{fig:prob}
 \vspace{-0.45cm}
\end{figure}

We evaluate the detection performance using probability of detection, $P_{\mathrm{D}}$, for a given probability of false alarm, which is given by
\begin{equation}
P_{\mathrm{D}} = \mathbb{E}[\varphi(E) \mid \text{target present}],
\end{equation}
where the thresholding function $\varphi(E)$ is defined in (\ref{eq:qD}). We perform Monte-Carlo simulations with 10,000 trials to compute $P_{\mathrm{D}}$ using the preamble-based detection technique at $P_{\mathrm{FA}}$ of $10^{-4}$, $10^{-6}$, and $10^{-8}$.
Fig.~\ref{fig:prob} shows the performance of the proposed detection algorithm as a function of probability of false alarm and the received SCNR. It indicates that $P_{\mathrm{D}}$ grows with increasing $P_{\mathrm{FA}}$. For a $P_{\mathrm{FA}}$ of $10^{-4}$, it is possible to achieve radar detection rates greater than 90\% above the received SCNR of -24.3 dB and for a $P_{\mathrm{FA}}$ of $10^{-6}$, it is possible to achieve $P_{\mathrm{D}} > $ 99.9\% for received SCNR $>$ -20.5 dB.
 
\begin{figure}[!t]
\centering
\includegraphics[scale=0.65]{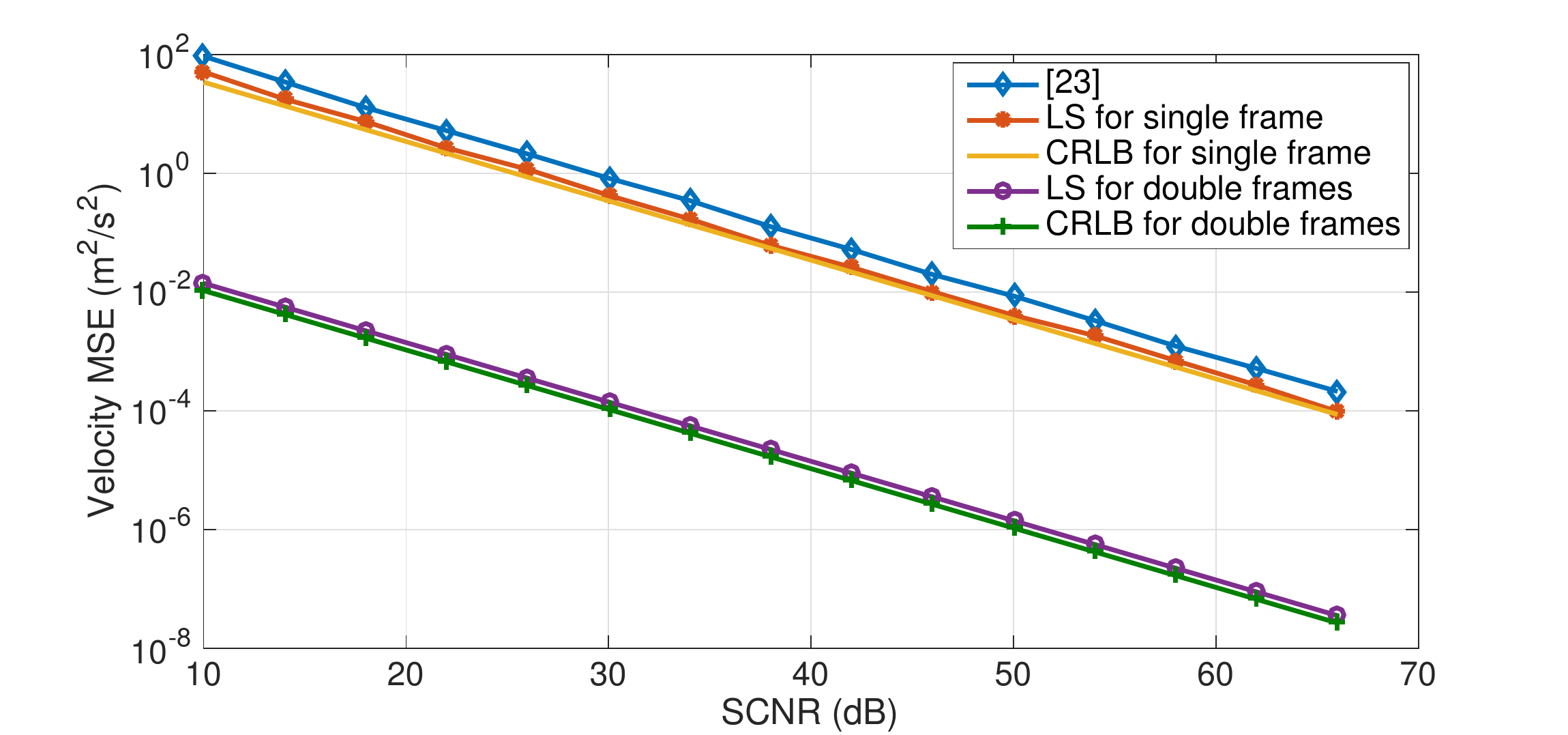}
 \caption{Estimated velocity MSE using the STF of a single and the preamble of double frames. The MSEs of proposed estimation techniques closely match with their CRLBs.}
\label{fig:vel}
 \vspace{-0.45cm}
\end{figure}

Fig.~\ref{fig:vel} shows the estimated velocity MSE using the STF of a single frame with $P=128 \times 16$ and $\ND=512$, and using the preamble of two frames with $P=128 \times 26$ and $\ND=41,285$. The estimated velocity MSEs increase linearly (in dB scale) with the SCNR. The velocity estimation using the LS-based algorithm in (\ref{eq:freqFinal}) is comparatively better than the one proposed in \cite{liuall}. The accuracy of the LS-based estimation techniques is very close to its CRLB bound. Using double frames, we achieve much better velocity estimation accuracy than using a single frame for all SCNR values. At low SCNR (less than 10 dB), however, even using double frames, we do not achieve the desired velocity accuracy of 0.1 m/s. 

\begin{figure}[!t] 
 \includegraphics[scale = 0.65]{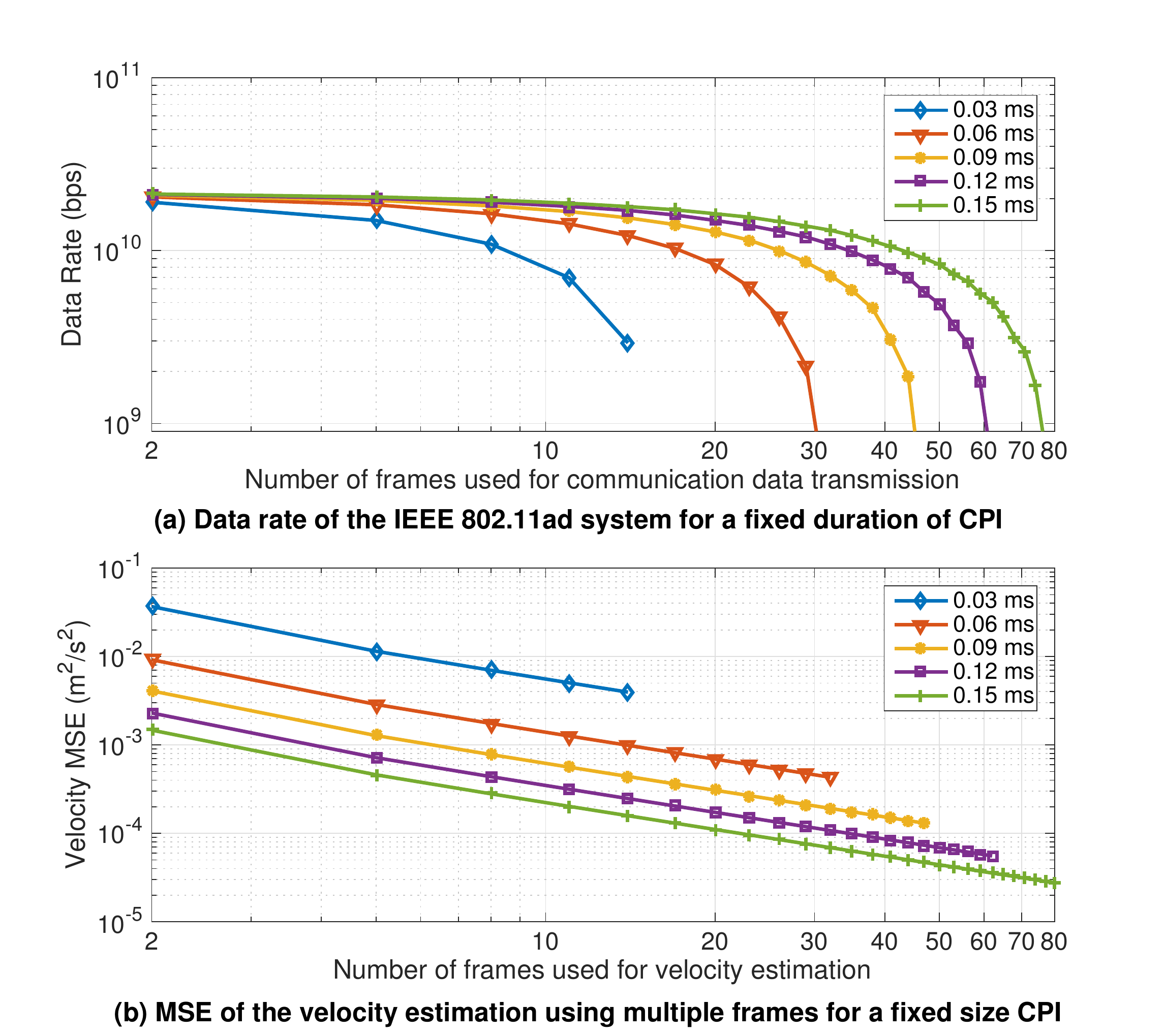}
\caption{Trade-off between communication data rate and velocity estimation MSE for a fixed size CPI. By increasing the duration of training symbols within a CPI, velocity estimation becomes more accurate with reduced data rate.}
\label{fig:mseData}
 \vspace{-0.45cm}
\end{figure}
This motivates us to exploit multiple frames as explained in Section \ref{Sec:ProRx}, which inherently increases the training sequence and frame duration to better estimate velocity using the LS-based method. The performance of this algorithm, however, depends on the number of frames during a CPI. To evaluate the dependence of velocity estimation on the number of frames within a CPI and investigate its simultaneous effect on the communication system, we consider the following data rate as the communication performance metric
\begin{equation}
R = \frac{M K_{\mathrm{CD}}\Ts}{T} \e{  \log_2 \left( 1+ \snr \right) },
\end{equation}
where $K_{\mathrm{CD}}$ is the total number of communication data symbols within a frame.

We have performed simulations over different CPI duration at 10 dB SCNR to investigate the trade-off between velocity estimation MSE and communication data rate, as shown in Fig.~\ref{fig:mseData}. For a fixed CPI duration, the number of frames in a CPI is varied from one to its maximum limit such that the number of symbols in each frame conforms to the IEEE 802.11ad SC PHY frame structure. We observe from the simulations that there is a trade-off between the communication data rate and the velocity estimation accuracy for a given CPI duration with a different number of frames. With an increase in the number of frames for a fixed CPI duration, the communication data rate degrades while enhancing the velocity estimation accuracy. We also observe that it is possible to simultaneously achieve Gbps communication data rate and cm/s-level accurate target velocity estimation for a CPI of 0.06 ms or more.

\begin{figure}[!t]
\centering
\includegraphics[scale = 0.6]{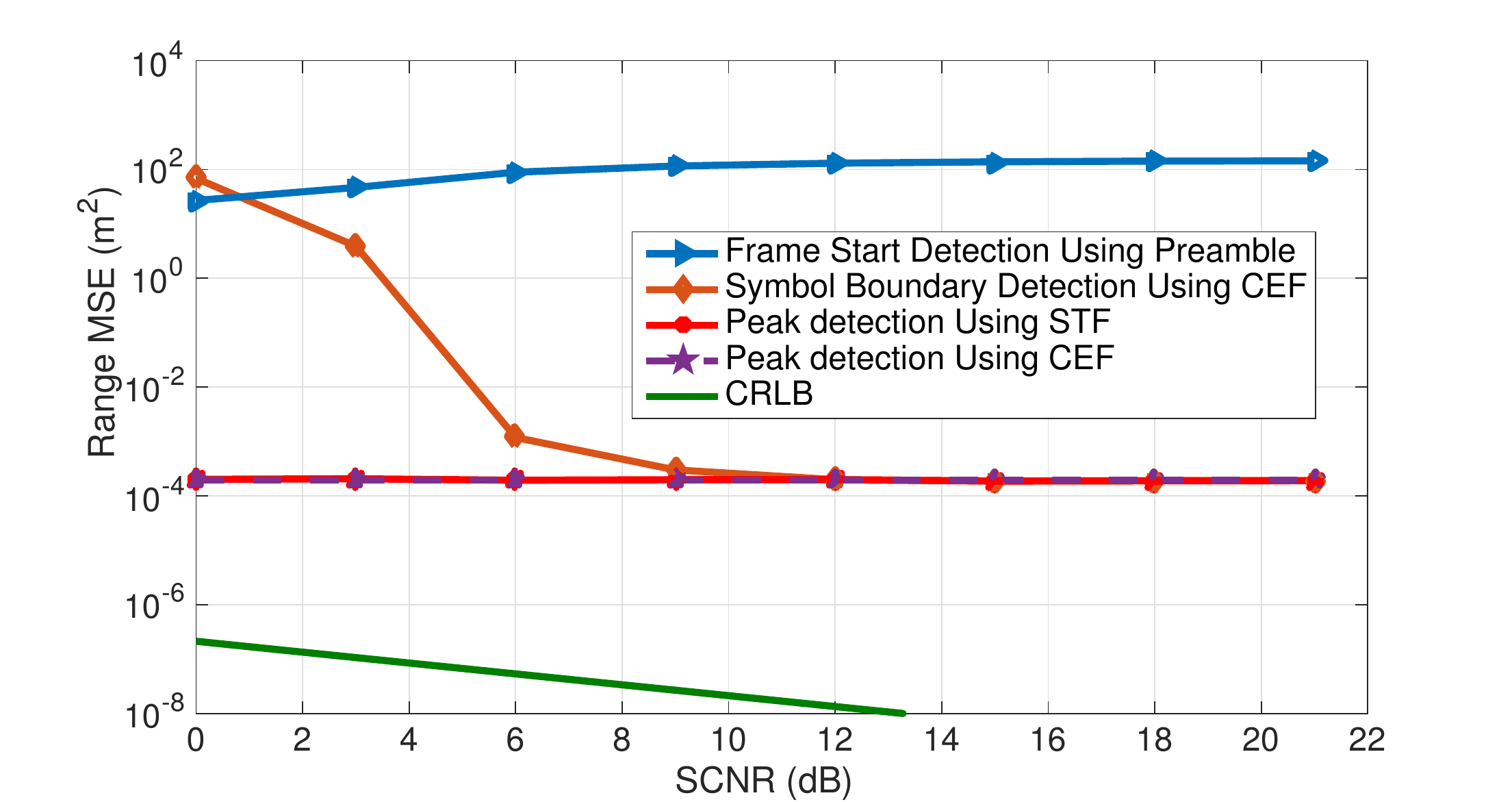}
 \caption{Estimated range MSE for coarse and fine range estimation algorithms using the preamble of a single frame.}
\label{fig:NAC_BAC_CE}
 \vspace{-0.45cm}
\end{figure}

In Fig.~\ref{fig:NAC_BAC_CE}, we compare the performance of various proposed range estimation algorithms and the CRLB using a single frame. The desired range MSE for automotive radars is 0.01 m$^2$ \cite{hasch2012millimeter}. For frame start detection using the STF, we chose a threshold of $\chi^2_{\mathrm{STF}} = 1/8$ because it reduces the complexity of the hardware implementation \cite{liuall}. We observe from Fig.~\ref{fig:NAC_BAC_CE} that the fine range estimation achieves better than the desired accuracy of 0.1 m using the STF/CEF peak detection for SCNR above 0 dB, and using the CEF symbol boundary detection for SCNR above 6 dB. The poor performance of range estimation using the CEF symbol boundary detection at low SCNR can be attributed to the fact the performance of the phase-based estimation is affected by Doppler shift. The figure also shows that the performance of the frame start detection using the preamble degrades due to a constant threshold $\chi_{\mathrm{STF}}$, which does not adapt to the increasing SCNR \cite{preyss2015digital}. This does not, however, degrade the performance of the fine range estimation technique. Indeed, the amplitude-based peak detection techniques using the STF/CEF achieve better than the desired automotive range accuracy of 0.1 m using a single frame without incorporating significant complexity. These peak detection techniques achieve range estimation MSEs quite close to the CRLB with a slight difference of less than 2 cm$^2$ between them. This difference is due to the fixed oversampling factor used in the symbol synchronization algorithm to estimate the fractional symbol delay, which can be easily improved using a higher oversampling factor or by using more complex symbol synchronization algorithm. 

\begin{figure}[!t]
\centering
\includegraphics[scale=0.6]{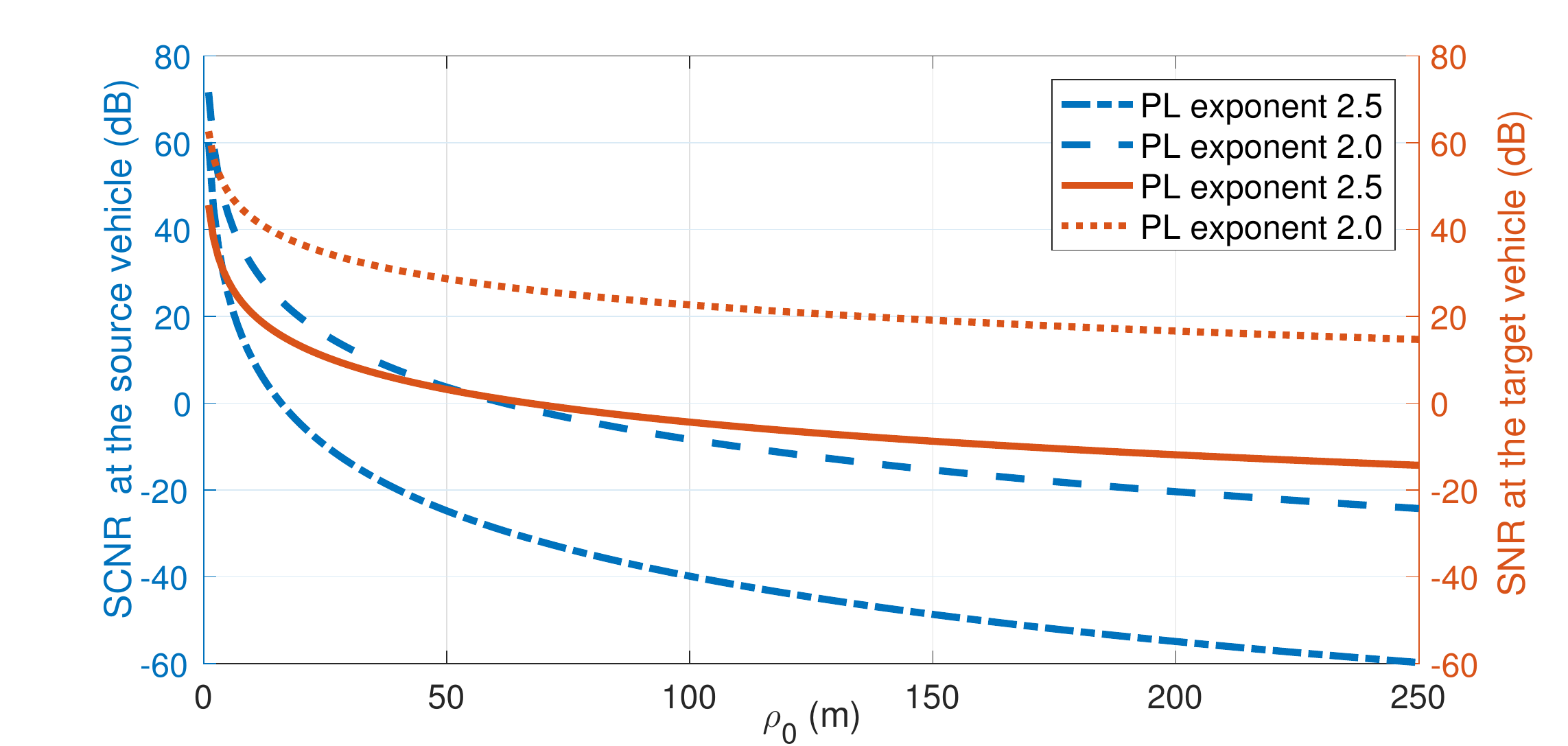}
 \caption{Received radar SCNR at the source vehicle and received communication SNR at the target vehicle as a function of distance between the source and target vehicles.}
\label{fig:snrVsdis}
 \vspace{-0.45cm}
\end{figure}
Fig.~\ref{fig:snrVsdis} shows the received radar SCNR at the source vehicle and the received communication SNR at the target vehicle decrease for a given TX EIRP of 43 dBm (maximum EIRP for USA \cite{ieee2012wireless}) with PL exponents of 2 (as used in automotive radar simulations, e.g., \cite{bazzi2012estimation}, and in mmWave communication propagation model, e.g., \cite{RapMacSam:Wideband-Millimeter-Wave-Propagation:15}) and 2.5 (as used in \cite{kim2013enabling} for link budget analysis of IEEE 802.11ad indoor applications) assuming noise figure of 6 dB \cite{kim2013enabling}. We also infer from Fig.~\ref{fig:snrVsdis} that for a given EIRP, one-way received communication SNR is higher than the two-way radar SCNR and both of them reduces with the increasing $\rho_\um$ due to decreasing large-scale channel gains $G_\comm$ and $G_\mr$. Based on Fig.~\ref{fig:prob} and Fig.~\ref{fig:snrVsdis}, we infer that for PL exponent of 2.0, the IEEE 802.11ad-based radar can detect very reliably with $P_{\mathrm{D}} > 99.9$\% and $P_{\mathrm{FA}} = 10^{-6}$ till 200 m, which is desirable for LRR \cite{Continental}. The span of $\rho_\um$ over which the proposed radar system can detect also overlaps with the range of $\rho_\um$ over which we can reliably estimate the cm-level accurate range and cm/s-level accurate velocity of the target vehicle within a CPI duration of less than 10 ms, which is desirable in LRR \cite{rohling2001waveform,hasch2012millimeter}.

\subsection{Multi-target Scenario}
\begin{figure}[!t]
\centering
\includegraphics[scale=0.65]{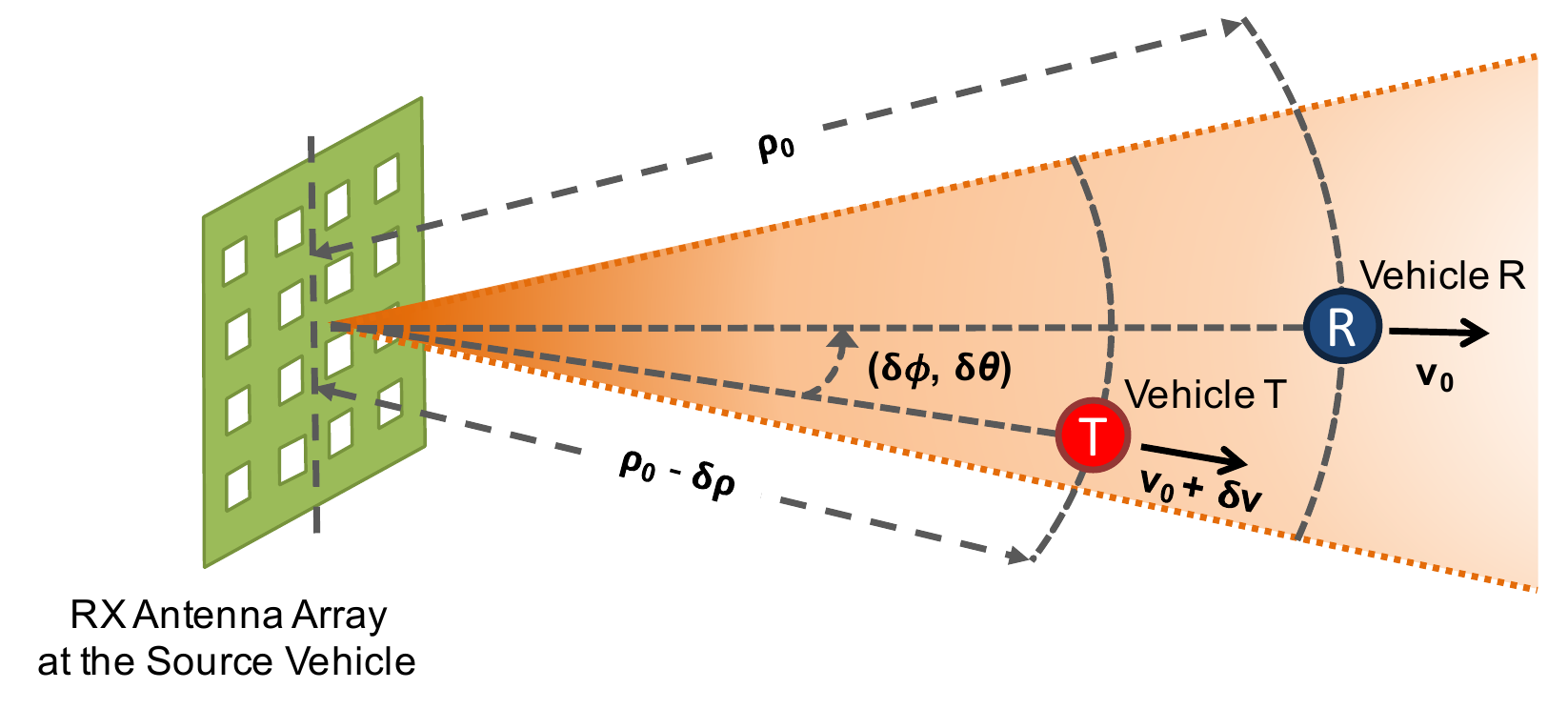}
 \caption{A multi-target scenario with two vehicles, namely vehicle R and vehicle T, within the mainlobe of the TX beam at the source vehicle. Both the vehicles are slightly separated in range, relative velocity, and direction w.r.t the source vehicle.}
\label{fig:multiScenario}
\vspace{-0.45cm}
\end{figure}
The performance of the IEEE 802.11ad-based joint communication-radar system is also evaluated for a multi-target vehicular scenario, where we use the doubly selective channel model as described in Section \ref{sec:System}. To demonstrate the range and velocity resolutions of the joint system, we consider a two-target vehicle scenario. In this scenario, one of the target vehicles is a recipient vehicle, say vehicle R. The second target vehicle is considered within the beamwidth of the source vehicle and is separated in range, relative velocity, and AoA/AoD as compared with the recipient vehicle by $\delta \rho$ and $\delta v$, and ($\delta \phi, \delta \theta$) respectively, as shown in Fig.~\ref{fig:multiScenario}. For vehicle R, we choose range as 14.32 m and velocity as 30 m/s, which falls in the typical operating span of LRR range and velocity specifications \cite{hasch2012millimeter}. We consider that the AoA/AoD corresponding to the vehicle R is $(90^{\circ},90^{\circ})$, which is likely to happen when the vehicle R is in the same lane as the source vehicle for applications such as cruise control. For vehicle T, we consider $\delta \rho =$ 4.26 m, $\delta v =$ 30 m/s, and ($\delta \phi, \delta \theta$) = $(100^{\circ},90^{\circ})$, which also falls in the typical span of LRR specifications \cite{hasch2012millimeter}.
\begin{figure}[!t]
\centering
\includegraphics[scale=0.6]{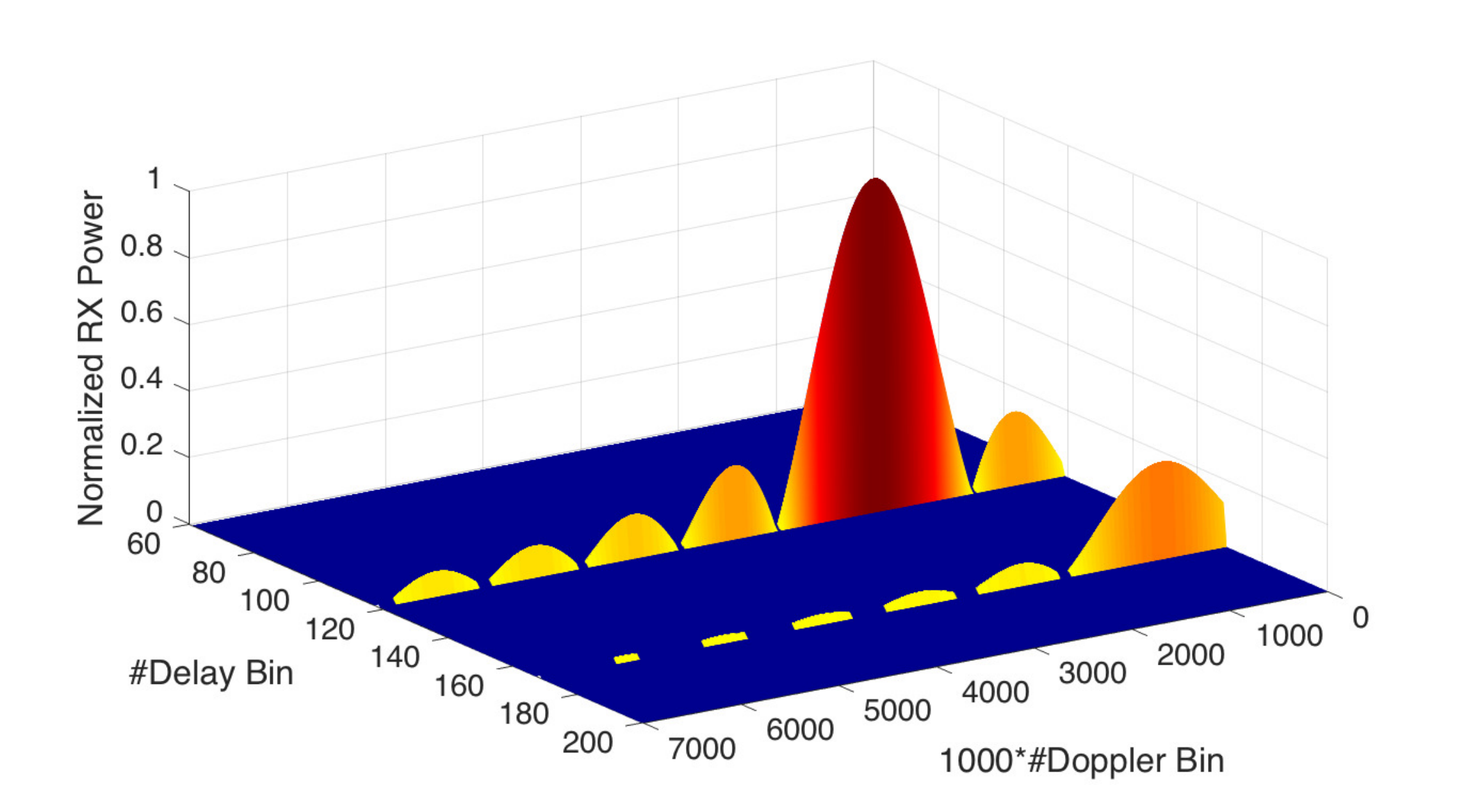}
 \caption{Mesh plot of the estimated delay-Doppler map. The plot shows two mainlobe peaks corresponding to the simulated vehicle R and vehicle T. Due to the broad mainlobe width in the Doppler domain velocity resolution is limited to around 35 m/s in this simulation.}
\label{fig:multiTargetMesh}
 \vspace{-0.45cm}
\end{figure}

\begin{figure}[!t]
\centering
\includegraphics[scale=0.6]{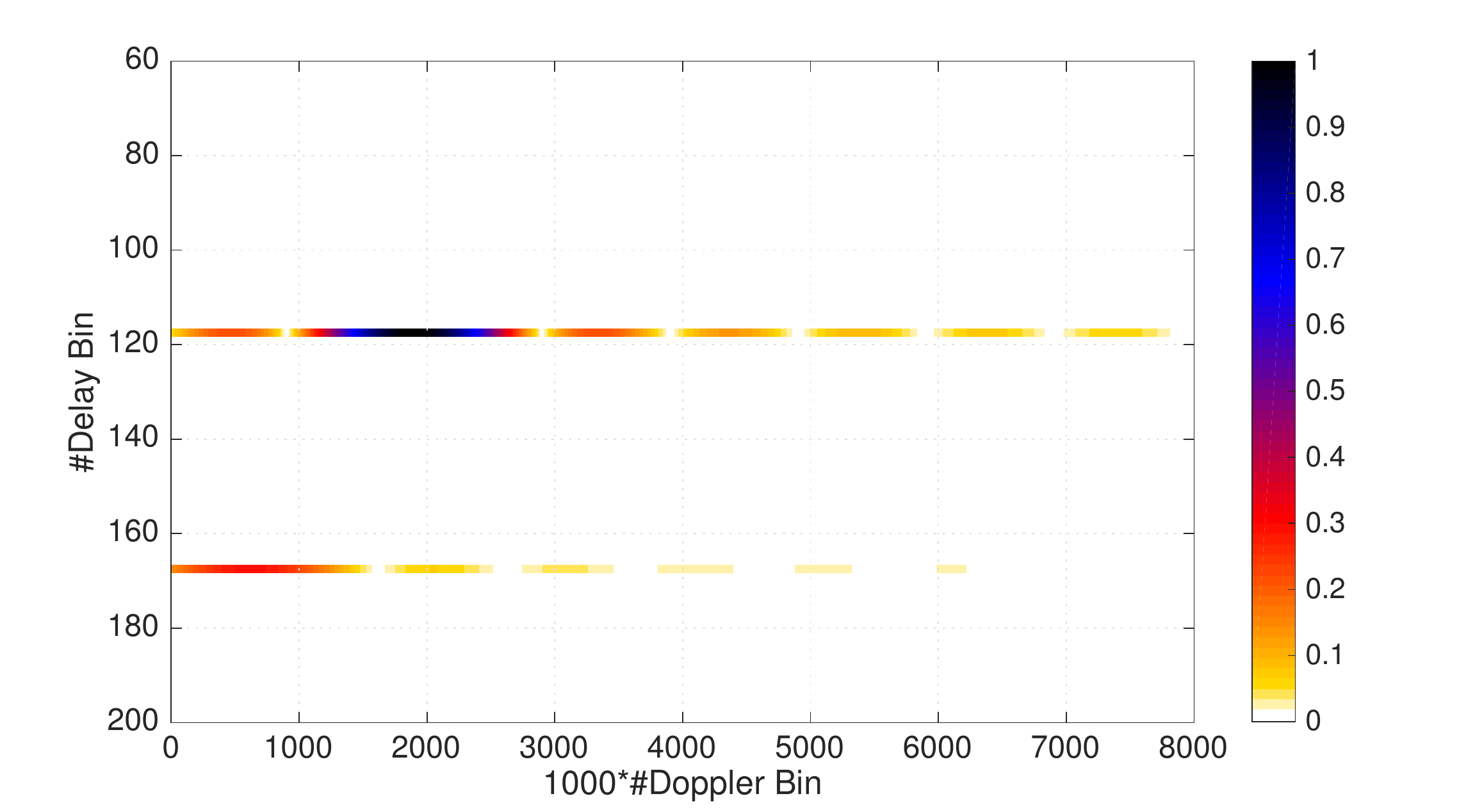}
 \caption{The estimated 2D-plot of the normalized delay-Doppler map shows that there are 2 dominant reflections present in the $118^\mathrm{th}$ and $168^\mathrm{th}$ delay bins and in the first and second Doppler bins.}
\label{fig:multiTargetImage}
 \vspace{-0.45cm}
\end{figure}

Figs. \ref{fig:multiTargetMesh} and \ref{fig:multiTargetImage} represent the 2D and three-dimensional (3D) plots of the estimated normalized delay-Doppler map with 10 frames in one CPI of 128000 samples, i.e., 0.072 ms duration. The normalization is done w.r.t the maximum received power at the source vehicle. The delay resolution is 0.07 m, and the Doppler resolution is 13750 KHz, calculated from Remark 3. In both figures, we plot the amplitude of the delay-Doppler map w.r.t the discrete-delay, $\ell$, and 1000 times interpolated discrete-Doppler, $d$, as described in Section \ref{sec:System}. The delay-Doppler map is interpolated in the Doppler (slow-time) dimension to visualize the resolution degradation effect due to the wide mainlobe and high sidelobes of the Doppler response in a given delay bin. These plots show that vehicle R and vehicle T responses are separated by 50 delay bins, corresponding to 4.26 m. They also show high sidelobes and wide mainlobe in the Doppler axis resulting in limited Doppler resolution. From the 3-dB mainlobe width along the Doppler dimension, we infer that the Doppler resolution is 13,750 KHz and the corresponding velocity resolution is 34.375 m/s. The velocity resolution, however, improved to less than 0.6 m/s, when we increased the CPI duration to more than 4.2 ms, as can also be seen from Remark 3. From the 3-dB mainlobe width along the delay dimension, we deduce that the range resolution is 7 cm. We can also infer from Fig.~\ref{fig:multiTargetMesh} that gain of vehicle R is less than vehicle T. This is because vehicle R is farther as compared to vehicle T from the source vehicle. This example illustrates the limit of velocity resolution, which is highly dependent on the duration of a CPI. At the same time, it also demonstrates the cm-level resolution and ultra-low sidelobes along the range dimension.

\section{Conclusions}
In this paper, we developed a mmWave automotive radar based on the IEEE 802.11ad standard to enable a joint mmWave vehicular communication-radar system. Our proposed radar receiver exploits the preamble structure (repeated Golay complementary sequences) of an IEEE 802.11ad SC PHY frame because of its perfect auto-correlation property at the zero-Doppler shift. We proposed different single- and multi-frame radar algorithms for the single- and multi-target radar detection as well as range and velocity estimation by leveraging standard WLAN receiver algorithms and classical pulse-Doppler radar algorithms. We evaluated the performance of these algorithms both analytically and by simulations. The target vehicle is detected very reliably at significantly low constant false alarm rate for SCNR above -20.5 dB. The range of the target vehicle is estimated with higher resolution and accuracy than the minimum requirement of the LRR specifications (0.5 m range resolution and 0.1 m range accuracy). The velocity estimation using single frame processing technique met the desired accuracy of 0.1 m/s above 45 dB SCNR, whereas using multi-frame processing technique achieved velocity accuracy of less than 0.1 m/s even at SCNR as low as -20.5 dB for a CPI of 4.2 ms. We also achieved the desired velocity resolution of less than 0.6 m/s for a CPI of 4.2 ms using multi-frame processing. Additionally, we showed that there is a trade-off between velocity estimation accuracy and communication data rate for a fixed CPI duration, while the desired velocity accuracy is simultaneously achieved with Gbps data rate for a CPI of 0.06 ms or more. These results indicate that IEEE 802.11ad-based joint communication-radar system is a promising option for next-generation automotive applications. Future work includes modifying the IEEE 802.11ad waveform that permits a trade-off between radar parameters’ estimation accuracy/resolution and communication data rate to adjust adaptively to the requirements imposed by different vehicular scenarios.

\appendices
\index{Appendices@\emph{Appendices}}%

\section{Cramer Rao Lower Bound for Velocity Estimation}
We derive the CRLB for the velocity estimation algorithm using the preamble of multiple IEEE 802.11ad SC PHY frames, as proposed in Section \ref{Sec:ProRx}. Consider the received training sequence $\mathbf{p} \in \mathbb{C}^{\Pt M \times 1}$ in (\ref{eq:pVelVector}), which can be expressed for $ 0 \leq i \leq P-1, ~ 0\leq m \leq M-1$ as
\begin{equation} \label{eq:velCRLBi}
p[k_i+mK] = r[k_i+mK] +z_m[k_i], 
\end{equation}
where $k_i$ is the sample number corresponding to the location of the training symbol in a frame, $r[k_i+mK] =\sqrt{\Es} h_\um s[k_i-\ell_\um] \me^{\jm  \omega_\um (k_i+mK) }$, $\omega_\um=2 \pi \nu_\um \Ts$, and $K$ is the total number of samples in a frame. Assuming perfect symbol synchronization, $z_m[k_i]$ is distributed as $\mathcal{N_C}(0,\sigma_{\cm\nm}^2)$. 

To calculate the CRLB for each element of the parameter vector $\Theta = [  \Es   \vert \hum \vert^2$, $\omega_\um, \angle \hum]$, we compute the Fisher information matrix $\mathbf{I}(\Theta)$. 
The CRLB corresponding to $\omega_\um = 2 \pi \nu_\um \Ts$ estimation, $\sigma^2_{\hat{\omega}}$, is the second diagonal element of $\mathbf{I}^{-1}(\Theta)$, $[\mathbf{I}^{-1}(\Theta)]_{2,2}$, i.e., $\sigma^2_{\hat{\omega}} =  [\mathbf{I}^{-1}(\Theta)]_{2,2}$ \cite{baronkin2001cramer,gansman1996single}, where
\begin{equation} \label{eq:CRLBvel}
 [\mathbf{I}^{-1}(\Theta)]_{2,2} \geq \xi \left[ \sum_{n=0}^{\Pt M-1} n^2 - \frac{1}{\Pt M}\left(\sum_{n=0}^{\Pt M-1} n \right)^2 \right]^{-1},
\end{equation}
with $\xi = {(\Pt M \Erm + 1})/( {2 \Pt M \left( {\Erm} \right) ^2})$ and  $n=k_i+mK$.

For $\Erm \gg (1/\Pt)$ and for consecutive samples in a single frame, i.e., $M=1$, the CRLB for estimation of the Doppler shift $\nu_\um$ in Hz, can be expressed as $\sigma^2_{\hat{\nu}} = { [\mathbf{I}^{-1}(\Theta)]_{2,2}}/{(4 \pi^2 \Ts^2)}$, which results in 
\begin{equation} 
\sigma^2_{\hat{\nu}} \geq \frac{6}{(2\pi)^2\Ts^2 \Pt(\Pt^2-1) \Erm} \approx \frac{6}{(2\pi)^2\Pt^3 \Ts^2 \Erm}. 
\end{equation} 


In case when $\mathbf{p}$ is composed of non-consecutive training sequence, i.e., $n=i+mK$, then for large number of frames, i.e., large $M$, we can simplify (\ref{eq:CRLBvel}) as
\begin{equation}
\sigma^2_{\hat{\nu}} \geq  \frac{6}{4\pi^2(M P^3 + M^3 P K^2)\Ts^2 \Erm}.
\end{equation}


\bibliographystyle{IEEEtran}
\bibliography{IEEEabrv,ref}

\begin{thebibliography}{10}
\providecommand{\url}[1]{#1}
\csname url@samestyle\endcsname
\providecommand{\newblock}{\relax}
\providecommand{\bibinfo}[2]{#2}
\providecommand{\BIBentrySTDinterwordspacing}{\spaceskip=0pt\relax}
\providecommand{\BIBentryALTinterwordstretchfactor}{4}
\providecommand{\BIBentryALTinterwordspacing}{\spaceskip=\fontdimen2\font plus
\BIBentryALTinterwordstretchfactor\fontdimen3\font minus
  \fontdimen4\font\relax}
\providecommand{\BIBforeignlanguage}[2]{{%
\expandafter\ifx\csname l@#1\endcsname\relax
\typeout{** WARNING: IEEEtran.bst: No hyphenation pattern has been}%
\typeout{** loaded for the language `#1'. Using the pattern for}%
\typeout{** the default language instead.}%
\else
\language=\csname l@#1\endcsname
\fi
#2}}
\providecommand{\BIBdecl}{\relax}
\BIBdecl

\bibitem{saponara2014highly}
S.~Saponara, M.~Greco, E.~Ragonese, G.~Palmisano, and B.~Neri, \emph{Highly
  Integrated Low Power Radars}.\hskip 1em plus 0.5em minus 0.4em\relax Artech
  House, 2014.

\bibitem{hasch2012millimeter}
J.~Hasch, E.~Topak, R.~Schnabel, T.~Zwick, R.~Weigel, and C.~Waldschmidt,
  ``{Millimeter-wave technology for automotive radar sensors in the 77 GHz
  frequency band},'' \emph{{IEEE} Transactions on Microwave Theory and
  Techniques}, vol.~60, no.~3, pp. 845--860, 2012.

\bibitem{papadimitratos2009vehicular}
P.~Papadimitratos, A.~La~Fortelle, K.~Evenssen, R.~Brignolo, and S.~Cosenza,
  ``Vehicular communication systems: Enabling technologies, applications, and
  future outlook on intelligent transportation,'' \emph{IEEE Communications
  Magazine}, vol.~47, no.~11, pp. 84--95, 2009.

\bibitem{kenney2011dedicated}
J.~B. Kenney, ``Dedicated short-range communications ({DSRC}) standards in the
  {United States},'' in \emph{Proceedings of the IEEE}, vol.~99, no.~7, 2011,
  pp. 1162--1182.

\bibitem{choi2016millimeter}
J.~Choi, V.~Va, N.~Gonzalez-Prelcic, R.~Daniels, C.~R. Bhat, and R.~W.
  Heath~Jr, ``Millimeter-wave vehicular communication to support massive
  automotive sensing,'' \emph{IEEE Communications Magazine}, vol.~54, no.~12,
  pp. 160--167, December 2016.

\bibitem{VaShiBan:Millimeter-Wave-Vehicular:16}
V.~Va, T.~Shimizu, G.~Bansal, R.~W. Heath~Jr \emph{et~al.}, ``Millimeter wave
  vehicular communications: A survey,'' \emph{Now: Foundations and Trends in
  Networking}, vol.~10, no.~1, 2016.

\bibitem{han2013joint}
L.~Han and K.~Wu, ``Joint wireless communication and radar sensing
  systems--state of the art and future prospects,'' \emph{IET Microwaves,
  Antennas \& Propagation}, vol.~7, no.~11, pp. 876--885, 2013.

\bibitem{saddik2007ultra}
G.~N. Saddik, R.~S. Singh, and E.~R. Brown, ``Ultra-wideband multifunctional
  communications/radar system,'' \emph{{IEEE} Transactions on Microwave Theory
  and Techniques}, vol.~55, no.~7, pp. 1431--1437, 2007.

\bibitem{sturm2011waveform}
C.~Sturm and W.~Wiesbeck, ``Waveform design and signal processing aspects for
  fusion of wireless communications and radar sensing,'' \emph{Proceedings of
  the IEEE}, vol.~99, no.~7, pp. 1236--1259, 2011.

\bibitem{berger2010signal}
C.~R. Berger, B.~Demissie, J.~Heckenbach, P.~Willett, and S.~Zhou, ``Signal
  processing for passive radar using {OFDM} waveforms,'' \emph{IEEE Journal of
  Selected Topics in Signal Processing}, vol.~4, no.~1, pp. 226--238, 2010.

\bibitem{reichardt2012demonstrating}
L.~Reichardt, C.~Sturm, F.~Grunhaupt, and T.~Zwick, ``Demonstrating the use of
  the {IEEE} 802.11p car-to-car communication standard for automotive radar,''
  in \emph{Proceedings of the 6th European Conference on Antennas and
  Propagation (EUCAP)}, 2012, pp. 1576--1580.

\bibitem{zhang200724ghz}
H.~Zhang, L.~Li, and K.~Wu, ``{24 GHz software-defined radar system for
  automotive applications},'' in \emph{Proceedings of the 10th European
  Conference on Wireless Technology}, October 2007, pp. 138--141.

\bibitem{han2010radar}
L.~Han and K.~Wu, ``Radar and radio data fusion platform for future intelligent
  transportation system,'' in \emph{Proceedings of the 7th European Radar
  Conference (EuRAD)}, September 2010, pp. 65--68.

\bibitem{Han:2016:ARC:2980100.2980106}
\BIBentryALTinterwordspacing
Y.~Han, E.~Ekici, H.~Kremo, and O.~Altintas, ``Automotive radar and
  communications sharing of the {79-GHz} band,'' in \emph{Proceedings of the
  First ACM International Workshop on Smart, Autonomous, and Connected
  Vehicular Systems and Services}, ser. CarSys '16.\hskip 1em plus 0.5em minus
  0.4em\relax New York, NY, USA: ACM, 2016, pp. 6--13. [Online]. Available:
  \url{http://doi.acm.org/10.1145/2980100.2980106}
\BIBentrySTDinterwordspacing

\bibitem{YehChoPre:Security-in-automotive-radar:16}
E.~Yeh, J.~Choi, N.~Prelcic, C.~Bhat, and R.~{Heath, Jr.}, ``Security in
  automotive radar and vehicular networks,'' \emph{accepted to Microwave
  Journal}, 2016.

\bibitem{rohling1996cfar}
H.~Rohling and R.~Mende, ``{OS CFAR} performance in a {77 GHz} radar sensor for
  car application,'' in \emph{Proceedings of 1996 CIE International Conference
  of Radar}, 1996, pp. 109--114.

\bibitem{rohling2001waveform}
H.~Rohling and M.-M. Meinecke, ``Waveform design principles for automotive
  radar systems,'' in \emph{CIE International Conference on Radar}, 2001, pp.
  1--4.

\bibitem{estep2014magnetic}
N.~A. Estep, D.~L. Sounas, J.~Soric, and A.~Al{\`u}, ``Magnetic-free
  non-reciprocity and isolation based on parametrically modulated
  coupled-resonator loops,'' \emph{Nature Physics}, vol.~10, no.~12, pp.
  923--927, 2014.

\bibitem{LiJosTao:Feasibility-study-on-full-duplex:14}
L.~Li, K.~Josiam, and R.~Taori, ``Feasibility study on full-duplex wireless
  millimeter-wave systems,'' in \emph{Proceedings of the International
  Conference on Acoustics, Speech, and Signal Processing (ICASSP)}, May 2014,
  pp. 2769--2773.

\bibitem{sabharwal2013band}
A.~Sabharwal, P.~Schniter, D.~Guo, D.~Bliss, S.~Rangarajan, and R.~Wichman,
  ``In-band full-duplex wireless: challenges and opportunities,'' \emph{IEEE
  Journal on Selected Areas in Communications}, vol.~32, no.~9, pp. 1637--1652,
  September 2013.

\bibitem{Continental}
Continental, ``{ARS 30X /-2 /-2C/-2T/-21 Long Range Radar},'' 2009.

\bibitem{richards2005fundamentals}
M.~A. Richards, \emph{Fundamentals of radar signal processing}.\hskip 1em plus
  0.5em minus 0.4em\relax Tata McGraw-Hill Education, 2005.

\bibitem{liuall}
W.-C. Liu, T.-C. Wei, Y.-S. Huang, C.-D. Chan, and S.-J. Jou, ``{All-Digital
  Synchronization for SC/OFDM Mode of IEEE 802.15.3c and IEEE 802.11ad},''
  \emph{{IEEE} Transactions on Circuits and Systems I: Regular Papers},
  vol.~62, no.~2, Feb. 2015.

\bibitem{preyss2015digital}
N.~A. Preyss and A.~Burg, ``{Digital Synchronization for Symbol-spaced IEEE
  802. 11ad Gigabit mmWave Systems},'' in \emph{Proceedings of the 22nd IEEE
  International Conference on Electronics, Circuits, and Systems (ICECS)}, no.
  EPFL-CONF-214535, 2015, pp. 637--640.

\bibitem{preeti2015}
P.~Kumari, N.~Gonzalez-Prelcic, and R.~W. Heath~Jr, ``I{nvestigating the IEEE
  802.11ad Standard for Millimeter Wave Automotive Radar},'' in
  \emph{Proceedings of the 82nd IEEE Vehicular Technology Conference},
  September 2015, pp. 3587--3591.

\bibitem{ieee2012wireless}
``{Wireless LAN Medium Access Control (MAC) and Physical Layer (PHY)
  Specifications. Amendment 3: Enhancements for Very High Throughput in the 60
  GHz Band},'' \emph{IEEE Std. 802.11ad}, 2012.

\bibitem{turyn1963ambiguity}
R.~Turyn, ``Ambiguity functions of complementary sequences (corresp.),''
  \emph{{IEEE} Transactions on Information Theory}, vol.~9, no.~1, pp. 46--47,
  1963.

\bibitem{bazzi2012estimation}
A.~Bazzi, C.~K{\"a}rnfelt, A.~Peden, T.~Chonavel, P.~Galaup, and F.~Bodereau,
  ``Estimation techniques and simulation platforms for {77 GHz FMCW ACC}
  radars,'' \emph{The European Physical Journal Applied Physics}, vol.~57, p.
  11001 (16 pp.), 2012.

\bibitem{RapMacSam:Wideband-Millimeter-Wave-Propagation:15}
T.~S. Rappaport, G.~R. MacCartney, M.~K. Samimi, and S.~Sun, ``Wideband
  millimeter-wave propagation measurements and channel models for future
  wireless communication system design,'' \emph{IEEE Transactions on
  Communications}, vol.~63, no.~9, pp. 3029--3056, Sept 2015.

\bibitem{roh2014millimeter}
W.~Roh, J.-Y. Seol, J.~Park, B.~Lee, J.~Lee, Y.~Kim, J.~Cho, K.~Cheun, and
  F.~Aryanfar, ``Millimeter-wave beamforming as an enabling technology for {5G}
  cellular communications: theoretical feasibility and prototype results,''
  \emph{IEEE Communications Magazine}, vol.~52, no.~2, pp. 106--113, 2014.

\bibitem{menzel2012antenna}
W.~Menzel and A.~Moebius, ``Antenna concepts for millimeter-wave automotive
  radar sensors,'' \emph{Proceedings of the IEEE}, vol. 100, no.~7, pp.
  2372--2379, 2012.

\bibitem{SonChoLov:Common-Codebook-Millimeter:16}
J.~Song, J.~Choi, and D.~J. Love, ``Common codebook millimeter wave beam
  design: Designing beams for both sounding and communication with uniform
  planar arrays,'' \emph{arXiv preprint arXiv:1606.05634}, 2016.

\bibitem{wilocity15}
\BIBentryALTinterwordspacing
{Wil6200: Second Generation WiGig and 802.11ad Multi-Gigabit Wireless Chipset}.
  [Online]. Available: \url{http://wilocity.com/resources/Wil6200-Brief.pdf}
\BIBentrySTDinterwordspacing

\bibitem{currie1987principles}
N.~C. Currie and C.~E. Brown, \emph{Principles and applications of
  millimeter-wave radar}.\hskip 1em plus 0.5em minus 0.4em\relax Artech House,
  1987.

\bibitem{saha201560}
S.~K. Saha, V.~V. Vira, A.~Garg, and D.~Koutsonikolas, ``{60 GHz Multi-Gigabit
  Indoor WLANs: Dream or Reality?}'' \emph{arXiv preprint arXiv:1509.04274v2},
  2016.

\bibitem{va2015impact}
V.~Va, J.~Choi, and R.~W. Heath~Jr, ``The impact of beamwidth on temporal
  channel variation in vehicular channels and its implications,''
  \emph{accepted to IEEE Transactions on Vehicular Technology, arXiv preprint
  arXiv:1511.02937}, Oct 2016.

\bibitem{sayeed2010wireless}
A.~Sayeed and T.~Sivanadyan, ``Wireless communication and sensing in multipath
  environments using multiantenna transceivers,'' \emph{Handbook on Array
  Processing and Sensor Networks}, 2010.

\bibitem{zhou2012efficient}
L.~Zhou and Y.~Ohashi, ``{Efficient codebook-based MIMO beamforming for
  millimeter-wave WLANs},'' in \emph{IEEE 23rd International Symposium on
  Personal Indoor and Mobile Radio Communications (PIMRC)}, 2012, pp.
  1885--1889.

\bibitem{MarIwaOht:First-Eigenmode-Transmission:16}
\BIBentryALTinterwordspacing
K.~Maruta, T.~Iwakuni, A.~Ohta, T.~Arai, Y.~Shirato, S.~Kurosaki, and
  M.~Iizuka, ``First eigenmode transmission by high efficient {CSI} estimation
  for multiuser massive {MIMO} using millimeter wave bands,'' \emph{Sensors
  (Basel, Switzerland)}, vol.~16, no.~7, p. 1051, 07 2016. [Online]. Available:
  \url{http://www.ncbi.nlm.nih.gov/pmc/articles/PMC4970098/}
\BIBentrySTDinterwordspacing

\bibitem{MacSunRap:Millimeter-Wave-Wireless:16}
\BIBentryALTinterwordspacing
G.~R. MacCartney, Jr., S.~Sun, T.~S. Rappaport, Y.~Xing, H.~Yan, J.~Koka,
  R.~Wang, and D.~Yu, ``Millimeter wave wireless communications: New results
  for rural connectivity,'' in \emph{Proceedings of the 5th Workshop on All
  Things Cellular: Operations, Applications and Challenges}, ser. ATC
  '16.\hskip 1em plus 0.5em minus 0.4em\relax New York, NY, USA: ACM, 2016, pp.
  31--36. [Online]. Available: \url{http://doi.acm.org/10.1145/2980055.2987353}
\BIBentrySTDinterwordspacing

\bibitem{AskEkm:Tracking-With-a-High-Resolution:15}
S.~A. Askeland and T.~Ekman, ``Tracking with a high-resolution {2D} spectral
  estimation based automotive radar,'' \emph{IEEE Transactions on Intelligent
  Transportation Systems}, vol.~16, no.~5, pp. 2418--2423, Oct 2015.

\bibitem{li2007mimo}
J.~Li and P.~Stoica, ``{MIMO radar with colocated antennas},'' \emph{{IEEE}
  Signal Processing Magazine}, vol.~24, no.~5, pp. 106--114, 2007.

\bibitem{LimJeoLee:Rejection-of-road-clutter:11}
H.~S. Lim, S.~H. Jeong, and K.~H. Lee, ``Rejection of road clutter using
  mean-variance method with {OS-CFAR} for automotive applications,'' in
  \emph{Proceedings of the International Conference on Electrical and Control
  Engineering (ICECE)}, Sept 2011, pp. 4886--4889.

\bibitem{Kle:Applications-of-space-time-adaptive:04}
R.~Klemm, \emph{Applications of space-time adaptive processing}.\hskip 1em plus
  0.5em minus 0.4em\relax IET, 2004, vol.~14.

\bibitem{liu2013digital}
W.-C. Liu, F.-C. Yeh, T.-C. Wei, C.-D. Chan, and S.-J. Jou, ``A digital
  {Golay-MPIC} time domain equalizer for {SC/OFDM} dual-modes at 60 {GHz}
  band,'' \emph{IEEE Transactions on Circuits and Systems I: Regular Papers},
  vol.~60, no.~10, pp. 2730--2739, 2013.

\bibitem{Roh:Some-radar-topics::06}
H.~Rohling, ``Some radar topics: waveform design, range {CFAR} and target
  recognition,'' in \emph{Advances in Sensing with Security
  Applications}.\hskip 1em plus 0.5em minus 0.4em\relax Springer, 2006, pp.
  293--322.

\bibitem{bo2015compressed}
G.~Bo, Z.~Changming, J.~Depeng, and Z.~Lieguang, ``{Compressed SNR-and-channel
  estimation for beam tracking in 60-GHz WLAN},'' \emph{Communications, China},
  vol.~12, no.~6, pp. 46--58, 2015.

\bibitem{moose1994technique}
P.~H. Moose, ``A technique for orthogonal frequency division multiplexing
  frequency offset correction,'' in \emph{IEEE Transactions on Communications},
  vol.~42, no.~10, 1994, pp. 2908--2914.

\bibitem{skolnik2008radar}
M.~I. Skolnik, Ed., \emph{Radar Handbook}, 3rd~ed.\hskip 1em plus 0.5em minus
  0.4em\relax McGraw Hill, 2008.

\bibitem{Zhu:2014:DOP:2639108.2639121}
\BIBentryALTinterwordspacing
Y.~Zhu, Z.~Zhang, Z.~Marzi, C.~Nelson, U.~Madhow, B.~Y. Zhao, and H.~Zheng,
  ``Demystifying {60GHz} outdoor picocells,'' in \emph{Proceedings of the 20th
  Annual International Conference on Mobile Computing and Networking}, ser.
  MobiCom '14.\hskip 1em plus 0.5em minus 0.4em\relax New York, NY, USA: ACM,
  2014, pp. 5--16. [Online]. Available:
  \url{http://doi.acm.org/10.1145/2639108.2639121}
\BIBentrySTDinterwordspacing

\bibitem{kim2013enabling}
J.~Kim and A.~F. Molisch, ``Enabling gigabit services for {IEEE}
  802.11ad-capable high-speed train networks,'' in \emph{IEEE Radio and
  Wireless Symposium (RWS)}, 2013, pp. 145--147.

\bibitem{baronkin2001cramer}
V.~M. Baronkin, Y.~V. Zakharov, and T.~C. Tozer, ``{Cramer-Rao} lower bound for
  frequency estimation in multipath {Rayleigh} fading channels,'' in
  \emph{Proceedings of International Conference on Acoustics, Speech, and
  Signal Processing (ICASSP'01)}, vol.~4, 2001, pp. 2557--2560.

\bibitem{gansman1996single}
J.~Gansman, J.~Krogmeier, and M.~Fitz, ``Single frequency estimation with
  non-uniform sampling,'' in \emph{Proceedings of 13th Asilomar Conference on
  Signals, Systems and Computers}, vol.~1, 1996, pp. 399--403.

\end{thebibliography}

\end{document}